\documentclass[prc,showpacs,floatfix]{revtex4}
\usepackage{bm}% bold math
\usepackage{dcolumn}
\usepackage{graphicx}% Include figure files
\usepackage{amsmath}
\usepackage{bbold}
\newcommand{\beq}{\begin{equation}}
\newcommand{\eeq}{\end{equation}}
\newcommand{\beqa}{\begin{eqnarray}}
\newcommand{\eeqa}{\end{eqnarray}}
\begin{document}
\title{
\hfill{\small {\bf MKPH-T-06-07}}\\
{\bf Incoherent single pion electroproduction on the deuteron with
polarization effects}}
\author{M. Tammam$^a$, A. Fix$^b$ and H. Arenh\"ovel$^b$}
\affiliation{
$^a$Physics Department, Al-Azhar University, Asiut, Egypt,\\
$^b$Institut f\"ur Kernphysik,
Johannes Gutenberg-Universit\"at Mainz, D-55099 Mainz, Germany}
\date{\today}
\begin{abstract}
Incoherent pion electroproduction on the deuteron is studied from
threshold up to the second resonance region with special emphasis on
the influence of final state interaction, in particular on
polarization observables. The elementary $\gamma N\rightarrow\pi N$
amplitude is taken from the MAID-2003 model. Final state interaction
is included by considering complete rescattering in the final $NN$ and
$\pi N$ subsystems. Their influence on the structure functions
governing the semi-exclusive differential cross section, where besides
the scattered electron only the produced pion is detected, is
investigated in detail. For charged pion-production the effect of 
$NN$-rescattering is moderate whereas $\pi N$-rescattering is almost
negligible, except very close to threshold. $NN$-rescattering appears
much stronger in neutral pion production for which the primary
mechanism is the elimination of a significant spurious coherent
contribution in the impulse approximation. Sizeable effects are also
found in some of the polarization structure functions for beam and/or
target polarizations.  
\end{abstract}

\maketitle

\section{Introduction}

The present paper is an extension of previous work on electromagnetic
single pion production on the deuteron~\cite{ArF05,FiA05} in which we
had considered the case of photoproduction. In the first
part~\cite{ArF05}, a thorough derivation of the formal expressions for
polarization observables in this reaction were presented. Then
in~\cite{FiA05} we had systematically investigated this process using
as realistic elementary pion production operator the MAID-2003
model~\cite{MAID} and included complete rescattering in the 
two-body $NN$- and $\pi N$-subsystems of the final state (FSI). Moderate
influences of the latter on total and semi-exclusive differential
cross sections were found in charged pion production, primarily from
$NN$-rescattering while $\pi N$-rescattering remained small. Much
larger effects were found in incoherent neutral pion production,
which, however, originated predominantly from the elimination of a
sizeable spurious coherent contribution to the incoherent process in
the impulse approximation (IA) where any final state interaction is
neglected. 

In view of the interest in this reaction with respect to (i) extracting 
on the one hand information on the elementary production on the
neutron in using the deuteron as an effective neutron target, and (ii)
studying the influence of the spectator nucleon, i.e.\ medium effects, it
appears natural to investigate the corresponding electroproduction
reaction taking advantage of the possibility to vary energy and
momentum transfer independently in the space like region. For example, 
it would be interesting to see whether some kinematic regions exist where
$\pi N$-rescattering becomes more important. Indeed, several studies
of the role of FSI and medium effects have already been undertaken in
the past, both experimentally~\cite{BrC71,GiB90,GaA01} as well as
theoretically~\cite{LoP94,HaL01,LeC04}. 

An early experiment by C.N.\ Brown et al.~\cite{BrC71} was designed to
study the isoscalar-isovector interference of the elementary amplitude
and the role of nuclear corrections by measuring on the one hand the
ratio of $\pi^-$ to $\pi^+$ production on the deuteron and on the
other hand the ratio of $\pi^+$ production on the deuteron to the one
on hydrogen. The 
forward-angle production of charged pions on the deuteron was measured
by R.\ Gilman et al.~\cite{GiB90} in order to investigate possible
influences of the spectator nucleon on the elementary production
amplitude. For the cross section ratio of $\pi^+$ production on the
deuteron to the one on the proton they found a significant deviation
from unity. Their conclusion was that there is evidence for a
modification of the elementary pion production process in the nuclear
medium. In a subsequent theoretical paper, Loucks et al.~\cite{LoP94}
obtained within a simple model for the elementary pion production
operator this ratio in fair agreement with experiment without invoking
medium modifications. The deviation from unity was traced back to the
strong final state interaction in the $^1S_0$-partial wave of the
outgoing neutrons in which the $^1S_0$-anti-bound state is
the dominant feature for low energies. In addition a strong 
dependence of the differential cross section on the tensor
polarization of an oriented deuteron was found which is a
manifestation of the non-spherical character of the deuteron via its
$D$-state component. 

A similar motivation with respect to possible medium modifications was
also behind a more recent experiment by D.~Gaskell et
al.~\cite{GaA01} which was triggered by the observation that
as long as the longitudinal current is dominated by the pion pole term
one could explore the nuclear pion field. Results were presented for
longitudinal charged pion production on $^1$H, $^2$H, and $^3$He
targets. The data, however, did not support any significant
modification of the elementary production process by the presence of
the spectator nucleons. At about the same time, Hafidi and
Lee~\cite{HaL01} published a theoretical study using a dynamical model
for the e.m.\ pion production operator, allowing besides final state
rescattering also to include intermediate baryon-baryon interactions,
i.e., two-body contributions to the e.m.\ interaction. However, the
latter turned out to be almost negligible in the near threshold
region. The same near-threshold region was also considered in another
more recent theoretical paper by Levchuck et al.~\cite{LeC04} using a 
unitary transformation method and restricting the e.m.\ current to the
lowest multipoles based on the Born contributions alone, i.e.\ leaving
out the contribution of the $\Delta$-resonance. The results
of~\cite{HaL01} were confirmed with respect to the influence of FSI
near the quasi-free peak but not for low missing mass. 

In the present work we would like to study more systematically
incoherent electroproduction of single pions on the deuteron using a
realistic 
elementary pion production operator with respect to the importance of
final state interactions in different energy regions, from threshold
through the $\Delta$-resonance up to the second resonance region. In
particular, the role of polarization degrees of freedom will be
explored. In the next section, we will briefly review the formal
aspects of this reaction, especially the definition of polarization
observables. The results will be presented and discussed in Sect.~4,
and we will close with a summary and an outlook. 

\section{Formalism}\label{formalism}
The basic formalism for electromagnetic single pion production on the
deuteron has been presented in detail for the case of photoproduction
in~\cite{ArF05}. Therefore, we review here only the most important
ingredients with due extensions to electroproduction according to the
additional contributions from charge and longitudinal current
components. 

\subsection{Kinematics}\label{kinematics}

The kinematics of pion electroproduction in the one-photon exchange
approximation is very similar to photoproduction in replacing
the real photon by a virtual one with longitudinal and transverse
polarizations 
\beq
\gamma^*(q)+d(p_d)\!\rightarrow\!
\pi(p_\pi)+N_1(p_1)+N_2(p_{2})\,,
\eeq
defining here the notation of the four-momenta of the participating
particles, i.e., $q=(q_0,\vec q\,)$ for the virtual photon,
$p_d=(E_d,\vec p_d)$ for the deuteron, 
$p_\pi=(E_\pi,\vec p_\pi)$ for the produced pion, and $p_i=(E_i,\vec
p_i)$ for the outgoing nucleons ($i=1,2$). The momentum of the virtual
photon is determined by the four-momentum transfer in the scattering
process, i.e.\ $q=k_e-k_{e'}$ denoting by $k_e=(E_e,\vec k_e)$ and
$k_{e'}=(E_{e'},\vec k_{e'})$ the momenta of incoming and scattered 
electrons, respectively. The electron kinematics will be considered in
the laboratory frame, while the evaluation of the reaction matrix will
be done in the center-of-momentum frame (c.m.) of virtual photon and
deuteron, i.e.\ all variables, which determine the reaction matrix,
refer to the c.m.\ frame if not indicated specifically otherwise.

As independent variables for the description of the final state
we choose in the c.m.\ frame the outgoing pion momentum $\vec
p_\pi=(p_\pi,\theta_\pi,\phi_\pi)$ and the 
spherical angles $\Omega_p=(\theta_p,\phi_p)$ of the relative momentum
$\vec p=(\vec p_1-\vec p_2)/2=(p,\Omega_p)$ of the two outgoing
nucleons having momenta $\vec p_1$ and $\vec p_2$. In conjunction with
the momentum of the virtual photon, the energies $E_i$ and momenta of the
outgoing nucleons are fixed, i.e.\ 
\beqa
E_{1/2}&=&\frac{1}{2}E_{12}\mp\frac{\vec p\cdot\vec p_\pi}{E_{12}}=
\frac{1}{2}E_{12}\mp\frac{pp_\pi}{E_{12}}\cos\theta_{p\pi}\,,\\
\vec p_{1/2}&=&-\frac{1}{2}\vec p_\pi \pm \vec p\,,
\eeqa
with $\theta_{p\pi}$ as angle between $\vec p$ and
$\vec p_\pi$ and $E_{12}=E_1+E_2=W-E_\pi$ as the total final $NN$ energy,
where 
\beq
W=q_0+\sqrt{M_d^2+q^2}=\sqrt{(2q_0^{lab}+M_d)M_d-Q^2}
\eeq
denotes the invariant total mass, $M_d$ the deuteron mass, and
$Q^2=q_\mu^2$. Furthermore, the square of the relative momentum is 
fixed by the independent variables and is given by
\beq
p^2=\frac{E_{12}^2(E_{12}^2-p_\pi^2-4M^2)}
{4(E_{12}^2-p_\pi^2\cos^2{\theta_{p\pi}})}\,,
\eeq
where the nucleon mass is denoted by $M$. The pion momentum is restricted to
$0\leq p_\pi \leq p_{\pi,max}$,
where the upper limit is given by
\beqa
p_{\pi,max}&=& \frac{1}{2W}\,\sqrt{((W-m_\pi)^2-4M^2)
((W+m_\pi)^2-4M^2)}\,.\label{p_pi_limit}
\eeqa

Of special interest is the quasi-free kinematics which is defined by
the condition that the spectator nucleon remains at rest in the lab
system, i.e.\ its final momentum is given by $p_s^{lab}=(M,\vec
0)$. In this case, the lab energy of the active final pion-nucleon
system is given by 
\beq\label{quasifreelab}
E_{\pi N}^{qf,lab}=\sqrt{m_\pi^2+(\vec p_\pi^{\,lab})^2}
+\sqrt{M^2+(\vec q^{\,lab}-\vec p_\pi^{\,lab})^2}=M_d-M+q_0^{lab}\,.
\eeq
For the semi-exclusive reaction, where besides the scattered electron
only the produced pion is measured, one can determine the quasi-free
lab pion energy $E_{\pi}^{qf, lab}$ from (\ref{quasifreelab}) and
finds 
\beqa\label{Epiquasifreelab}
E_{\pi}^{qf, lab}(\theta_\pi^{lab})&=&
\frac{1}{2((E_{\pi
N}^{qf,lab})^2-(q^{\,lab})^2\cos^2\theta_\pi^{lab})}
\nonumber\\
&&\times\Big(C_{qf}^{lab}E_{\pi N}^{qf,lab}\pm q^{\,lab}\cos\theta_\pi^{lab}
\sqrt{(C_{qf}^{lab})^2-4m_\pi^2((E_{\pi N}^{qf,lab})^2-(q^{\,lab})^2
\cos^2\theta_\pi^{lab})}\Big),
\eeqa
where we have introduced
\beq
C_{qf}^{lab}=(E_{\pi N}^{qf,lab})^2+m_\pi^2-M^2-(q^{\,lab})^2
=(M_{\pi N}^{qf})^2+m_\pi^2-M^2\,, 
\eeq
with $M_{\pi N}^{qf}$ as invariant mass of the active quasi-free $\pi
N$-system. In (\ref{Epiquasifreelab}) the ``plus''-sign should be taken
for $0\leq \theta_\pi \leq \pi$, otherwise the ``minus''-sign. The
corresponding quasi-free missing mass $M_x^{qf}$ is given by 
\beq
M_x^{qf}=\sqrt{2M(M_d+q_0^{lab}-E_{\pi}^{qf, lab})}\,.
\eeq
In the c.m.\ system the quasi-free condition for the final spectator
momentum reads $p_s=(\sqrt{M^2+q^2/4},\vec
q/2)$. The corresponding expressions 
for the quasi-free pion energy and missing mass are 
\beqa\label{Epiquasifreecm}
E_{\pi}^{qf}(\theta_\pi)&=&
\frac{1}{2((E_{\pi N}^{qf})^2-q^2\cos^2\theta_\pi)}
\nonumber\\
&&\times
\Big(C_{qf}E_{\pi N}^{qf}\pm q\cos\theta_\pi
\sqrt{(C_{qf})^2-4m_\pi^2((E_{\pi N}^{qf})^2-q^2
\cos^2\theta_\pi)}\Big)\,,\\
M_x^{qf}&=&\sqrt{2W(W-2E_{\pi}^{qf})+m_\pi^2}\,,
\eeqa
with
\beq
E_{\pi N}^{qf}=W-\sqrt{M^2+q^2/4}\,,\quad\mbox{and } 
C_{qf}=(E_{\pi N}^{qf})^2+m_\pi^2-M^2-q^2\,.
\eeq

\begin{figure}[htb]
\includegraphics[scale=.6]{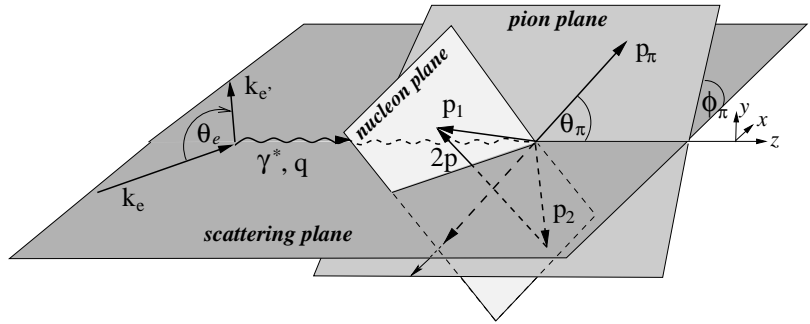}
\caption{Kinematics of single pion electroproduction on the deuteron.}
\label{fig_elpion_kinematik}
\end{figure}

As coordinate system we choose a right-handed orientation with
$z$-axis along the photon momentum $\vec{q}$ and $y$-axis
perpendicular to the scattering plane along $\vec k_e\times\vec k_{e'}$. We
distinguish in general three planes: (i) the scattering plane spanned by
the incoming and scattered electron momenta, (ii) the
pion plane, spanned by the photon and pion momenta, which intersects the
scattering plane along the $z$-axis with an angle $\phi_\pi$, and (iii) the
nucleon plane spanned by the momenta of the two outgoing nucleons 
intersecting the pion plane along the total momentum of the two
nucleons. This is illustrated in Fig.~\ref{fig_elpion_kinematik}. 

\subsection{The $T$-matrix}

As in photoproduction, all observables are determined by the
$T$-matrix elements of the 
electromagnetic pion production current $J_{\gamma\pi}$ between the
initial deuteron and the final $\pi NN$ states
\beq
T_{s m_s, \mu m_d}= -^{(-)}\langle \vec p_1\,\vec p_2\,s
m_s,\,\vec p_\pi\,| J_{\gamma\pi,\,\mu}(0)|
\vec p_d\,1 m_d\rangle\,,\label{Tmatrix}
\eeq
where $s$ and $m_s$ denote the total spin and its projection on the
relative momentum $\vec p$ of the outgoing two nucleons, and $m_d$ 
correspondingly the deuteron spin projection on the $z$-axis as 
quantization axis. In the expression on the rhs of (\ref{Tmatrix})
non-covariant normalization for the initial deuteron and the final
$\pi NN$-states is adopted. As already mentioned, all kinematic
quantities related to the $T$-matrix refer to the $\gamma^*$-$d$ c.m.\
system. Furthermore, the e.m.\ current is taken in
the normalization of the MAID-2003 model. Because of current
conservation, one can eliminate either the charge or the longitudinal
current component. In this work we have eliminated the longitudinal
current component. 

Separating the c.m.-motion and making a multipole
expansion of the current, its general form is given by 
\begin{eqnarray}
T_{s m_s \mu m_d}(W,Q^2,p_\pi,\Omega_\pi,\Omega_p)&=& -
^{(-)}\langle \vec p\, s m_s,\,\vec p_\pi\,|J_{\gamma\pi,\,\mu}(\vec
q\,)|1m_d\rangle\nonumber\\ 
&=& \sqrt{2\pi}\sum_{L} i^L\hat L 
^{(-)}\langle \vec p\, s m_s,\,\vec p_\pi\,
|{\cal O}^{\mu L}_\mu|1m_d\rangle\,, 
\end{eqnarray}
with $\mu\in\{0,\pm 1\}$ enumerating the spherical
current components with the provision that $J_{\gamma\pi,\,0}$ is
identified with the charge density. Furthermore, we use the
notation $\hat L=\sqrt{2L+1}$, and the symbol ${\cal O}^{\mu L}_M$
comprises charge ($C_M^L$) and transverse multipoles ($E_M^L$ and $M_M^L$) 
\begin{eqnarray}
{\cal O}^{\mu L}_M&=& \delta_{\mu 0}C_M^L +\delta_{|\mu| 1}(E_M^L +\mu
M_M^L)\,. 
\end{eqnarray}
Introducing a partial wave decomposition of the
final states, one finds 
\begin{eqnarray}\label{small_t}
T_{s m_s \mu m_d}(W,Q^2,p_\pi,\Omega_\pi,\Omega_p)&=& 
e^{i(\mu+m_d-m_s)\phi_\pi}
t_{s m_s \mu m_d}(W,Q^2,p_\pi,\theta_\pi,\theta_p,\phi_{p\pi})\,,
\end{eqnarray}
where the small $t$-matrix depends besides $W$, $Q^2$ and $p_\pi$ only
on $\theta_\pi$, $\theta_p$, and the relative azimuthal angle
$\phi_{p\pi}=\phi_p-\phi_\pi$. Explicitly one has (for details we
refer to~\cite{ArF05}) 
\begin{eqnarray}
t_{s m_s \mu m_d}(W,Q^2,p_\pi,\theta_\pi,\theta_p,\phi_{p\pi})&=&
\frac{1}{2\,\sqrt{2\pi}}
\sum_{L l_p j_p m_p l_\pi m_\pi J M_J}i^L\,\hat L\,\hat J\,\hat l_\pi
\,\hat l_p\,\hat j_p\,(-)^{J+l_p+j_p-s+m_s-l_\pi}\nonumber\\
&&\times\left(\begin{array}{ccc} 
l_p & s & j_p \cr 0 & m_s & -m_s\cr
\end{array}\right)
\left(\begin{array}{ccc} 
j_p & l_\pi & J \cr m_p & m_\pi & -M_J\cr
\end{array}\right)
\left(\begin{array}{ccc} 
J & L & 1 \cr -M_J & \mu & m_d\cr
\end{array}\right)\nonumber\\
&&\times\langle p \,p_\pi ((l_ps)j_p l_\pi)J||{\cal O}^{\mu L}||1\rangle
d^{j_p}_{m_s,m_p}(-\theta_p)\,d^{l_\pi}_{0,m_\pi}(-\theta_\pi)\,
e^{i(m_p-m_s)\phi_{p\pi}}\,.\label{smallt}
\end{eqnarray}
We had shown in~\cite{ArF05} that, if parity is conserved, the following
symmetry relation holds for $\mu=\pm 1$ 
\begin{eqnarray}\label{symmetry}
t_{s -m_s -\mu -m_d}(W,Q^2,p_\pi,\theta_\pi,\theta_p,\phi_{p\pi})&=&
(-)^{s+m_s+\mu+m_d}
t_{s m_s \mu m_d}(W,Q^2,p_\pi,\theta_\pi,\theta_p,-\phi_{p\pi})\,.
\end{eqnarray}
One should note the sign change of $\phi_{p\pi}$ on the right-hand side.
It is easy to see that this relation holds also for $\mu=0$, noting
that the parity selection rules for charge transitions are the
same as for electric ones. As pointed out in~\cite{ArF05}, all
observables can be expressed in terms of the small $t$-matrix
elements. 

In the present work we include as e.m.\ current the elementary
one-body pion production current of MAID-2003 and consider as FSI the
rescattering contributions in the final $NN$- and $\pi N$-subsystems. 
Thus as in~\cite{FiA05} we split the $T$-matrix into the impulse
approximation (IA) $T^{IA}$, where final state interaction effects are
neglected, and the rescattering contribution $T^{NN}$ and $T^{\pi N}$
of the two-body $NN$- and $\pi N$-subsystems, respectively,
\beq
T_{s m_s \mu m_d}=T_{s m_s \mu m_d}^{IA}+T_{s m_s \mu
m_d}^{NN}+T_{s m_s \mu m_d}^{\pi N}\,.
\eeq
For the IA contribution, where the final state is described by a
plane wave, antisymmetrized with respect to the two outgoing nucleons,
one has 
\beqa
T_{sm_s\mu m_d}^{IA}&=&\langle \vec p\, s m_s,\,\vec p_\pi\,|\,
\Big[j_{\gamma\pi,\mu}(1)+j_{\gamma\pi,\mu}(2)\Big]|\,1\,m_d\rangle
\nonumber\\
&=&\sqrt{2}\sum_{m_s^{\prime}}\Big(\langle sm_s\,|\,\langle \vec p_1|
j_{\gamma\pi,\mu}(W_{\gamma N_1},Q^2)|\vec p_d-\vec p_2\rangle
\phi_{m_s^{\prime}m_d}(\frac{1}{2}\vec p_d-\vec{p}_2) 
|\,1\,m_s'\rangle-(1\leftrightarrow 2)\Big)
\,,
\eeqa
where $j_{\gamma\pi,\mu}$ denotes the elementary pion photoproduction
operator of the MAID-2003 model, $W_{\gamma N_1}$ the 
invariant energy of the $\gamma N_1$ system, 
$\vec p_{1/2}=(\vec q+\vec p_d-\vec p_\pi)/2\pm\,\vec p$. Furthermore, 
$\phi_{m_sm_d}(\vec{p}\,)$ is related to the internal deuteron wave
function in momentum space by 
\beq
\langle \vec{p},
1m_s|1m_s\rangle^{(d)}=\phi_{m_sm_d}(\vec{p}\,)=\sum_{L=0,2}\sum_{m_L}i^L
(Lm_L\,1m_s|1m_d)u_L(p)\,Y_{Lm_L}(\hat{p})\,, \label{dwave}
\eeq
normalized to unity. The two rescattering contributions have a similar
structure 
\beqa
T_{s m_s \mu m_d}^{NN}&=&\langle\vec p\, s m_s,\,\vec p_\pi\,|\,
T_{NN}G_{NN}[j_{\gamma\pi,\mu}(W_{\gamma N_1},Q^2)
+j_{\gamma\pi,\mu}(W_{\gamma N_2},Q^2)]|\,1\,m_d\rangle\,,\\
T_{s m_s \mu m_d}^{\pi N}&=&\langle\vec p\, s m_s,\,\vec p_\pi\,|\,
T_{\pi N}G_{\pi N}[j_{\gamma\pi,\mu}(W_{\gamma N_1},Q^2)
+j_{\gamma\pi,\mu}(W_{\gamma N_2},Q^2)]|\,1\,m_d\rangle\,,
\eeqa
where $T_{NN}$ and $T_{\pi N}$ denote respectively the $NN$ and $\pi
N$ scattering matrices and $G_{NN}$ and $G_{\pi N}$ the corresponding
free two-body propagators. 

\subsection{The differential cross section including polarization observables}
\label{diff_cross_elpion}

The standard expression of the differential cross section for
electroproduction of pions on the deuteron in the one-photon-exchange
approximation is
\beqa
\frac{d^8\sigma}{dE_{e'} d\Omega_{e'} dp_\pi d\Omega_\pi d\Omega_p}&=&
\frac{\alpha_{qed}}{Q^4}\,\frac{k_{e'}}{k_e}\,
c(W,Q^2,p_\pi,\Omega_\pi,\Omega_p)\,tr(T^\dagger T\rho_i)\,,
\label{diffcross_elpion}
\eeqa
where $\alpha_{qed}$ denotes the e.m.\ fine structure constant, $T$
the reaction matrix, and $\rho_i$ the initial state 
density matrix for the spin degrees of virtual photon and deuteron. 
The trace refers to all spin degrees of freedom of initial and final
states. Furthermore, a kinematic phase space factor is denoted by 
\beq
c(W,Q^2,p_\pi,\theta_\pi,\theta_p,\phi_{p\pi})=\frac{M^2p^2p_\pi^2}
{4(2\pi)^4E_\pi(E_{12}p+\frac{1}{2}p_\pi(E_1-E_2)\cos{\theta_{p\pi}})}\,.
\eeq
The density matrix $\rho_i$ in (\ref{diffcross_elpion}) is a
direct product of the density matrices $\rho^{\gamma^*}$ of the virtual
photon and $\rho^d$ of the deuteron 
\begin{equation}
\rho_i=\rho^{\gamma^*}\otimes\rho^d \,.
\end{equation}
One can now proceed in complete analogy on the one hand to deuteron
electrodisintegration~\cite{ArL05} with respect to the virtual
photon and deuteron density matrices and on the other hand to pion
photoproduction with respect to the properties of the reaction
matrix. 

The virtual photon density matrix is determined by the electron
kinematics and separates into an unpolarized and a polarized part
\begin{equation}
\rho^{\gamma^*}_{\lambda \lambda'}=
\rho^0_{\lambda \lambda'}+h\rho'_{\lambda \lambda'}\,,
\end{equation}
where $|h|$ denotes the degree of longitudinal electron polarization,
and $\rho^0$ and $\rho'$ are given in terms of independent components 
$\rho_\alpha$ and $\rho^{\prime}_\alpha$ $(\alpha\in\{L,\,T,\,LT,\,TT\})$ 
according to the various combinations of longitudinal and transverse 
polarizations. Its specific form depends on whether one eliminates the
charge or the longitudinal current. In the latter case, as used in
this work, one has~\cite{ArL05}
\begin{subequations}\label{rhos}
\beqa
\rho^0_{\lambda \lambda'}&=&\sum_{\alpha\in\{L,\,T,\,LT,\,TT\}}
\delta^\alpha_{\lambda \lambda'}\rho_\alpha\,,\\
\rho'_{\lambda \lambda'}&=&\sum_{\alpha\in\{L,\,T,\,LT,\,TT\}}
\delta^{\prime\,\alpha}_{\lambda \lambda'}\rho'_\alpha\,,
\eeqa
\end{subequations}
with
\begin{equation}
\begin{array}{ll}
\delta^L_{\lambda \lambda'}=\delta_{\lambda \lambda'}\delta_{\lambda 0}\,,
&\delta^{LT}_{\lambda \lambda'}=\lambda'\delta_{\lambda 0}+
\lambda\delta_{\lambda' 0}\,,\cr
&\cr
\delta^T_{\lambda \lambda'}=\delta_{\lambda \lambda'}|\lambda|\,,
&\delta^{TT}_{\lambda \lambda'}=\delta_{\lambda,\, -\lambda'}|\lambda|
\,,\cr
&\cr
\delta^{\prime\,L}_{\lambda \lambda'}=0\,,
&\delta^{\prime\,LT}_{\lambda \lambda'}=|\lambda'|\delta_{\lambda 0}+
|\lambda|\delta_{\lambda' 0}\,,\cr
&\cr
\delta^{\prime\,T}_{\lambda \lambda'}=\delta_{\lambda \lambda'}
\lambda\,,
&\delta^{\prime\,TT}_{\lambda \lambda'}=0.\cr
\end{array}
\end{equation}
The independent components $\rho_\alpha$ and $\rho^{\prime}_\alpha$ 
are given by the well-known expressions~\cite{ArL05} (note $Q^2=-q_\nu^2>0$)
\begin{eqnarray}
\begin{array}{ll}
 \rho_L=\rho_{00}^0=\beta^2 Q^2\frac{\xi^2}{2\eta} 
\,,\quad& \rho_T=\rho_{11}^0
  =\frac{1}{2}Q^2\,\Big(1+\frac{\xi}{2 \eta} \Big) \,,\cr
&\cr
 \rho_{LT}=\rho_{01}^0=\beta Q^2 \frac{\xi}{\eta}\,
 \sqrt{\frac{\eta+ \xi}{8}}
\, ,& \rho_{TT}=\rho_{-11}^0=-Q^2\frac{\xi}{4 \eta} \,,\cr
&\cr
 \rho_{LT}^{\prime}=\rho_{01}^{\prime}=
 \frac{1}{2}\,\beta\frac{Q^2}{\sqrt{2\eta}}\,\xi \,,\quad&
 \rho_T^{\prime}=\rho_{11}^{\prime}=
  \frac{1}{2}Q^2\, \sqrt{\frac{\eta+\xi}{\eta}} \, ,\cr
\end{array}
\end{eqnarray}
with
\begin{equation}
\beta = \frac{ q^{\,\mathrm{lab}}}{q^{\,c}},\,\,\,\,\,
\xi = \frac{Q^2}{({q}^{\,\mathrm{lab}})^{\,2}},\,\,\,\,\,
\eta = {\rm tan}^2(\frac{\theta_e^{\mathrm{lab}}}{2})\;,\label{betaxieta}
\end{equation}
where $\beta$ expresses the boost from the lab system to the frame 
in which the 
hadronic current is evaluated and $\vec q^{\,c}$ denotes the momentum
transfer in this frame. Here it is the c.m.\ system, and one has $\vec
q^{\,c}=\vec q$. As a sideremark, we would like to mention
the simple relation to another often used parametrization of 
the virtual photon density matrix in terms of the quantities 
$v_{\alpha^{(\prime)}}$ of Ref.~\cite{DoS86} (for $\beta=1$ for the
lab frame, i.e.\ $q^c=q^{\mathrm{lab}}$) 
\begin{eqnarray}
\rho_\alpha^{(\prime)} &=& \frac{Q^2}{2\eta}\, v_{\alpha^{(\prime)}}\,,
\end{eqnarray}
where $\alpha \in\{ L,\, T,\, LT,\, TT\}$.

Assuming that the deuteron density matrix is diagonal 
with respect to an orientation axis $\vec d$ having spherical angles 
$(\theta_d,\phi_d)$ with respect to the coordinate system associated with 
the scattering plane in the lab frame, one has with respect to
$\vec d$ as quantization axis 
\begin{equation}
\rho_{m_d\, {m_d}'}^d=\frac{1}{\sqrt{3}}(-)^{1-m_d}
\sum_{I\,M}\hat{I}
\left( 
\begin{matrix}
1&1&I \cr m_d'&-m_d&M \cr
\end{matrix} \right) P_I^d
e^{-iM\phi_d}d^I_{M0}(\theta_d)\,. \label{rhoda}
\end{equation}
This means, 
the deuteron target is characterized by four parameters, namely the 
vector and tensor polarization parameters $P_1^d$ and $P_2^d$, respectively,
and by the orientation angles $\theta_d$ and $\phi_d$. The orientation 
parameters are related to the probabilities $\{p_m\}$ for finding a
deuteron spin projection $m$ on the orientation axis by
\beqa
P_I^d&=&\delta_{I 0} + \sqrt{\frac{3}{2}}(p_1-p_{-1})\,\delta_{I 1} 
+\frac{1}{\sqrt{2}}\,(1-3\,p_0)\,\delta_{I 2}\,.\label{rhodpar}
\eeqa
If one chooses the c.m.\ frame as reference frame as in the present
work, one should note that the deuteron density matrix undergoes no
change in the transformation from the lab to the c.m. system, since
the boost to the c.m.\ system is collinear with the deuteron
quantization axis~\cite{Rob74}. 

Following the same steps as in~\cite{ArF05}, one finds for the general
eight-fold differential cross section for single pion electroproduction 
with longitudinally polarized electrons 
\beqa
\frac{d^8\sigma}{dE_{e'} d\Omega_{e'} dp_\pi d\Omega_\pi d\Omega_p}&=&
\frac{\alpha_{qed}}{Q^4}\,\frac{k_{e'}}{k_e}\,\sum_{I=0}^2P^d_I\Big[
\rho_L\sum_{M\ge 0}d^I_{M0}(\theta_d)
\Big(\tau_L^{IM}\cos(M\phi_{\pi d})+\sigma_L^{IM}\sin(M\phi_{\pi d})\Big)\nonumber\\
&&\hspace*{-2cm}+\rho_T\sum_{M\ge 0}d^I_{M0}(\theta_d)
\Big(\tau_T^{IM}\cos(M\phi_{\pi d})+\sigma_T^{IM}\sin(M\phi_{\pi d})\Big)%\nonumber\\
+\rho_{LT}\sum_{M=-I}^Md^I_{M0}(\theta_d)
\Big(\tau_{LT}^{IM}\cos\phi_{M}+\sigma_{LT}^{IM}\sin\phi_{M}\Big)\nonumber\\
&&\hspace*{-2cm}+\rho_{TT}\sum_{M=-I}^Md^I_{M0}(\theta_d)
\Big(\tau_{TT}^{IM}\cos\psi_{M}+\sigma_{TT}^{IM}\sin\psi_{M}\Big)%\nonumber\\
+h\rho_T'\sum_{M\ge 0}d^I_{M0}(\theta_d)
\Big(\tau_{T}^{\prime IM}\cos(M\phi_{\pi d})+\sigma_{T}^{\prime IM}
\sin(M\phi_{\pi d})\Big)\nonumber\\
&&\hspace*{-2cm}+h\rho_{LT}'\sum_{M=-I}^Md^I_{M0}(\theta_d)
\Big(\tau_{LT}^{\prime IM}\cos\phi_{M}+\sigma_{LT}^{\prime IM}\sin\phi_{M}\Big)
\Big]\,,\label{exclusive_diff}
\eeqa
with 
\beq
\phi_{\pi d}=\phi_\pi-\phi_{d},\,\, \phi_{M}=M\phi_{\pi d}-\phi_{\pi},
\mbox{ and } \psi_{M}=M\phi_{\pi d}-2\phi_{\pi}\,.\label{phi}
\eeq
Again we would like to remind the reader that the electron kinematics
refer to the lab system while the final state kinematic variables and
the structure functions refer in this work to the c.m.\ system, which
we have chosen for the evaluation of the $T$-matrix. However, we would
like to point out that the expression for the differential cross
section in (\ref{exclusive_diff}) holds in general irrespective of
which frame of reference collinear with $\vec q$ is chosen for the
evaluation. Obviously, then the final state variables $p_\pi,\,
\theta_\pi,\, \theta_p$, and $\phi_{p\pi}$ refer to this frame.

The various exclusive structure functions $\tau_{\alpha}^{(\prime)IM}$ and
$\sigma_{\alpha}^{(\prime)IM}$ constitute the polarization observables which
determine beam, target and beam-target asymmetries. The structure
functions are defined by, not indicating all the kinematic variables
$W,\,Q^2,\,p_\pi,\, \theta_\pi,\, \theta_p$, and $\phi_{p\pi}$ on which they 
depend,  
\beqa
(\tau/\sigma)_{L}^{IM}&=&\pm\frac{1}{1+\delta_{M0}}\,
\Re e/\Im m \,u_{IM}^{00}\,,\quad M\ge 0\,,\label{tauL}\\
(\tau/\sigma)_{T}^{IM}&=&\pm\frac{1}{1+\delta_{M0}}\,
\Re e/\Im m \,(u_{IM}^{11}+u_{IM}^{-1-1})\,,\quad M\ge 0\,,\label{tauT}\\
(\tau/\sigma)_{LT}^{IM}&=&\pm\,\Re e/\Im m \,(u_{IM}^{10}-
u_{IM}^{0-1})\,,\label{tauLT}\\
(\tau/\sigma)_{TT}^{IM}&=&\pm\,\Re e/\Im m
\,u_{IM}^{1-1}\,,\label{tauTT}\\
(\tau/\sigma)_{T}^{\prime IM}&=&\pm\frac{1}{1+\delta_{M0}}\,
\Re e/\Im m \,(u_{IM}^{11}-u_{IM}^{-1-1})\,,\quad
M\ge 0\,,\label{tauTp}\\
(\tau/\sigma)_{LT}^{\prime IM}&=&\pm\,\Re e/\Im m \,(u_{IM}^{10}+
u_{IM}^{0-1})\,,\label{tauLTp}
\eeqa
in terms of the quantities introduced in~\cite{ArF05}
\beqa
u_{IM}^{\mu'\mu}(W,Q^2,p_\pi,\theta_\pi, \theta_p, \phi_{p\pi})&=& 
c(W,Q^2,p_\pi,\theta_\pi,\theta_p,\phi_{p\pi})\,
\frac{\hat I}{\sqrt{3}}\,\sum_{m_d m_d'}(-)^{1-m_d}
\left( 
\begin{matrix}
1&1&I \cr m_d'&-m_d&M \cr
\end{matrix} \right)\nonumber\\
&& \times
\sum_{s m_s}t^*_{s m_s \mu' m_d'}(W,Q^2,p_\pi,\theta_\pi, \theta_p,
\phi_{p\pi})\,t_{s m_s \mu m_d}(W,Q^2,p_\pi,\theta_\pi, \theta_p,
\phi_{p\pi})\,.\label{uim}
\eeqa
In~\cite{ArF05} we have shown that they behave under complex
conjugation as  
\beq\label{complc}
(u_{IM}^{\mu'\mu}(W,Q^2,p_\pi,\theta_\pi, \theta_p, \phi_{p\pi}))^*=
(-)^M\,u_{I-M}^{\mu\mu'}(W,Q^2,p_\pi,\theta_\pi, \theta_p, \phi_{p\pi})\,.
\eeq
From this property follows in particular that the $u_{I0}^{\mu\mu}$ are
real. Furthermore, the $u_{IM}^{\mu'\mu}$ possess the symmetry property
\beq
u_{IM}^{-\mu'-\mu}(W,Q^2,p_\pi,\theta_\pi, \theta_p, \phi_{p\pi})=
(-)^{I+M+\mu'+\mu}\,u_{I-M}^{\mu'\mu}(W,Q^2,p_\pi,\theta_\pi, \theta_p,
-\phi_{p\pi})\,, \label{sym_uim}
\eeq
which yields in combination with (\ref{complc})
\beq\label{complca}
u_{IM}^{-\mu'-\mu}(W,Q^2,p_\pi,\theta_\pi,\theta_p,\phi_{p\pi})=
(-)^{I+\mu'+\mu}\,(u_{IM}^{\mu\mu'}(W,Q^2,p_\pi,\theta_\pi, \theta_p,
-\phi_{p\pi}))^*\,.
\eeq
At the photon point, one finds the following equivalencies to the
corresponding quantities in pion photoproduction defined in~\cite{ArF05}
\beq
(\tau/\sigma)_{T}^{IM} =
\frac{q_0 W}{\pi^2 E_d} (\tau/\sigma)^{0}_{IM}\,,\quad
(\tau/\sigma)_{TT}^{IM} =
\frac{q_0 W}{\pi^2 E_d} (\tau/\sigma)^{l}_{IM}\,,\quad
(\tau/\sigma)_{T}^{\prime IM} =
\frac{q_0 W}{\pi^2 E_d} (\tau/\sigma)^{c}_{IM}\,,\label{equivalences}
\eeq
where obviously the kinematic variables on the right hand sides should
refer to the same reference frame then on the left hand sides.

For the semi-exclusive reaction $\vec d(\vec e,e'\pi)NN$, where besides
the scattered electron only the produced pion is detected, the basic
quantities are obtained from the
$u_{IM}^{\mu'\mu}(p_\pi,\,\theta_\pi,\, \theta_p,\, \phi_{p\pi})$ in
(\ref{uim}) by integration over $d\Omega_p$. Thus we introduce
\beqa
U_{IM}^{\mu'\mu}(W,Q^2,p_\pi, \theta_\pi)&=&\int d\,\Omega_{p}\,
u_{IM}^{\mu'\mu}(W,Q^2,p_\pi,\,\theta_\pi,\,\theta_p,\,\phi_{p\pi})
\nonumber\\&=& 
\frac{\hat I}{\sqrt{3}}\,\int
d\,\Omega_{p}\,c(W,Q^2,p_\pi,\theta_\pi,\theta_p,\phi_{p\pi})\, 
\sum_{m_d m_d'}(-)^{1-m_d}
\left( 
\begin{matrix}
1&1&I \cr m_d'&-m_d&M \cr
\end{matrix} \right)\nonumber\\
&& \times\sum_{s m_s}
(t_{s m_s \mu' m_d'}(W,Q^2,p_\pi,\theta_\pi,\theta_p,\phi_{p\pi}))^*
\,t_{s m_s \mu m_d}(W,Q^2,p_\pi, \theta_\pi,\theta_p,\phi_{p\pi})\,.
\eeqa
From the properties of (\ref{complc}) and (\ref{sym_uim}) one obtains
corresponding properties 
\beqa
(U_{IM}^{\mu'\mu})^*= (-)^MU_{I-M}^{\mu\mu'}\quad\mbox{and}\quad
U_{IM}^{-\mu'-\mu}=(-)^{I+M+\mu+\mu'}(U_{IM}^{\mu\mu'})^*\,.\label{VIM}
\eeqa
Combining them leads to
\beq
U_{IM}^{-\mu'-\mu}=(-)^{I+\mu+\mu'}(U_{IM}^{\mu\mu'})^*\,.
\eeq
An important consequence of this latter property is that, according to
(\ref{tauL}) through (\ref{tauLTp}), the following integrated
structure functions vanish for $\alpha\in\{L,T,LT,TT\}$
\beqa
\int d\,\Omega_{p}\,\tau^{1M}_\alpha&=&0\,\,\mbox{ and }
\,\,\int d\,\Omega_{p}\,\sigma^{\prime 1M}_{\alpha}=0\,,\\
\int d\,\Omega_{p}\,\sigma^{IM}_\alpha&=&0\,\,\mbox{ and }
\int d\,\Omega_{p}\,\tau^{\prime IM}_{\alpha}=0\,\,\mbox{ for }I=0,2\,.
\eeqa
The remaining semi-exclusive structure functions govern the six-fold
semi-exclusive differential cross section for which we find as final
form 
\beqa
\frac{d^6\sigma}{dE_{e'} d\Omega_{e'} dp_\pi d\Omega_\pi}&=&
\frac{\alpha_{qed}}{Q^4}\,\frac{k_{e'}}{k_e}\,\sum_{I=0}^2P^d_I\Big[
\rho_L\sum_{M\ge 0}d^I_{M0}(\theta_d)
\widetilde f_L^{IM}\cos(M\phi_{\pi d}-\delta_{I1}\frac{\pi}{2})\nonumber\\
&&+\rho_T\sum_{M\ge 0}d^I_{M0}(\theta_d)
\widetilde f_T^{IM}\cos(M\phi_{\pi
d}-\delta_{I1}\frac{\pi}{2})%\nonumber\\&&
+\rho_{LT}\sum_{M=-I}^Md^I_{M0}(\theta_d)
\widetilde f_{LT}^{IM}\cos(\phi_{M}-\delta_{I1}\frac{\pi}{2})\nonumber\\
&&+\rho_{TT}\sum_{M=-I}^Md^I_{M0}(\theta_d)
\widetilde
f_{TT}^{IM}\cos(\psi_{M}-\delta_{I1}\frac{\pi}{2})%\nonumber\\&&
+h\rho_T'\sum_{M\ge 0}d^I_{M0}(\theta_d)
\widetilde f_{T}^{\prime IM}\sin(M\phi_{\pi d}
+\delta_{I1}\frac{\pi}{2})\nonumber\\
&&+h\rho_{LT}'\sum_{M=-I}^Md^I_{M0}(\theta_d)
\widetilde f_{LT}^{\prime IM}\sin(\phi_{M}+\delta_{I1}\frac{\pi}{2})
\Big]\,,\label{diffcross_semi_inclusive}
\eeqa
where the angles $\phi_{\pi d}$, $\phi_{M}$, and $\psi_{M}$ are 
defined in (\ref{phi}). The semi-exclusive structure functions 
$\widetilde f_\alpha^{(\prime)IM}$ are given by 
\beqa\label{strufu}
\renewcommand{\arraystretch}{1.8}
\begin{array}{ll}
\widetilde f_L^{IM}=\frac{i^{\delta_{I1}}}{1+\delta_{M0}}\,
U^{00}_{IM}\,,&
\widetilde f_T^{IM}=\frac{2}{1+\delta_{M0}}\,
\Re e(i^{\delta_{I1}}U^{11}_{IM})\,,\cr
\widetilde f_{LT}^{IM}=2\,
\Re e(i^{\delta_{I1}}U^{10}_{IM})\,,&
\widetilde f_{TT}^{IM}=
\Re e(i^{\delta_{I1}}U^{1-1}_{IM})\,,\cr
\widetilde f_{LT}^{\prime\,IM}=2\,
\Im m(i^{\delta_{I1}}U^{10}_{IM})\,,&
\widetilde f_{T}^{\prime\,IM}=\frac{2}{1+\delta_{M0}}\,
\Im m(i^{\delta_{I1}}U^{11}_{IM})\,.\cr
\end{array}
\eeqa
They depend on $W$, $Q^2$, $p_\pi$, and $\theta_\pi$.
Because $U^{11}_{I0}$ is real according to (\ref{VIM}),
the structure functions $\widetilde f^{10}_T$ and $\widetilde
f^{\prime 20}_{T}$ vanish identically. We would like to point out that
for forward and backward pion emission, i.e.\ for $\theta_\pi=0$ and
$\pi$, the following structure functions have to vanish
\beq
\widetilde f^{IM}_{L}=0\,\,\mbox{ and }\,\,
\widetilde f^{(\prime)IM}_{T}=0\,\,\mbox{ for}\,\,M\neq 0,
\,\,\widetilde f^{(\prime)IM}_{LT}=0\,\,\mbox{ for}\,\,M\neq 1,
\mbox{ and }\,\, T^{IM}_{TT}=0\,\,\mbox{ for}\,\,M\neq 2\,,
\label{asym0}
\eeq
because in that case the differential cross section cannot depend on
$\phi_\pi$, since at $\theta_\pi=0$ or $\pi$ the azimuthal angle
$\phi_\pi$ is undefined or arbitrary. This feature can also be shown
by straightforward evaluation of $U_{IM}^{\mu'\mu}$ using the explicit
representation of the $t$-matrix in (\ref{smallt}) as shown
in~\cite{ArF05}. 

In case that only the direction of the outgoing pion is measured 
and not its momentum, the corresponding differential cross section
$d^5\sigma/(dE_{e'} d\Omega_{e'} d\Omega_\pi)$ is given by an expression
formally analogous to (\ref{diffcross_semi_inclusive}) where only the above
structure functions are integrated over the pion momentum, i.e., by the
replacement 
\beq
\widetilde f_{\alpha}^{(\prime)IM}(W,Q^2,p_\pi,\theta_\pi)
\rightarrow f_{\alpha}^{(\prime)IM}(W,Q^2,\theta_\pi)=
\int_{0}^{p_{\pi,max}} dp_\pi
\widetilde f_{\alpha}^{(\prime)IM}(p_\pi,\theta_\pi)\,
\eeq
for $\alpha\in\{L,T,LT,TT\}$. The upper integration limit is
listed in (\ref{p_pi_limit}).

The general totally inclusive cross section with respect to the
hdronic final state $d(e,e')\pi NN$ is obtained
from~(\ref{diffcross_semi_inclusive}) by integration over both $p_\pi$
and $\Omega_\pi$ resulting in 
\beqa
\frac{d^3\sigma}{dE_{e'} d\Omega_{e'}}&=&
\frac{\alpha_{qed}}{Q^4}\,\frac{k_{e'}}{k_e}
\Big[\rho_{L} F_{L}^{00}+\rho_T F_{T}^{00} %\nonumber\\&&
+P^d_1(h\rho_T' F_{T}^{\prime 10}d^1_{00}(\theta_d)+
[\rho_{LT} F_{LT}^{11}+h\rho_{LT}' F_{LT}^{\prime 11}]
d^1_{10}(\theta_d))\cos\phi_d \nonumber\\
&&+P^d_2([\rho_{L} F_{L}^{20}+\rho_T F_{T}^{20}]d^2_{00}(\theta_d)
+\rho_{LT} F_{LT}^{21}d^2_{10}(\theta_d)\cos\phi_d
+\rho_{TT} F_{TT}^{22}d^2_{20}(\theta_d)\cos (2\phi_d))\Big]
\eeqa
in terms of various form factors
\beqa
F_{\alpha}^{(\prime)IM}(W,Q^2)&=&\int d\Omega_\pi
\int_{0}^{p_{\pi,\,max}}dp_\pi\,\widetilde
f_{\alpha}^{(\prime)IM}(W,Q^2,p_\pi,\theta_\pi) \nonumber\\
&=&2\pi
\int d(\cos\theta_\pi) \,
f_{\alpha}^{(\prime)IM}(W,Q^2,\theta_\pi)\,.
\eeqa
At the photon point one obtains from (\ref{equivalences}) the
following relations to the contributions to the total photoproduction
cross section as listed in eq.~(83) of~\cite{ArF05}
\beq
F_T^{00}=\frac{q_0 W}{\pi^2 E_d}\sigma_0\,,\quad
F_{T}^{20}=\frac{q_0 W}{\pi^2 E_d}\overline T_{20}^{\,0}\,,\quad
F_{T}^{\prime 10}=\frac{q_0 W}{\pi^2 E_d}\overline T_{10}^{\,c}\,,\quad
F_{TT}^{22}=\frac{q_0 W}{\pi^2 E_d}\overline T_{22}^{\,l}\,.
\eeq

Finally, we would like to point out that for coherent electroproduction
of $\pi^0$ on the deuteron formally the same expression as in
(\ref{diffcross_semi_inclusive}) holds with structure functions, which
are defined in analogy to (\ref{strufu}) with the replacement
\beqa
U_{IM}^{\mu'\mu}&\rightarrow& c(W,Q^2,\theta_\pi)\,
\frac{\hat I}{\sqrt{3}}\,\sum_{m_d m_d'}(-)^{1-m_d}
\left( 
\begin{matrix}
1&1&I \cr m_d'&-m_d&M \cr
\end{matrix} \right)
\sum_{m_d''}t^*_{m_d'' \mu' m_d'}(W,Q^2,\theta_\pi)
\,t_{m_d'' \mu m_d}(W,Q^2,\theta_\pi)\,,
\eeqa
where $c(W,Q^2,\theta_\pi)$ denotes a kinematic factor.

\section{Results and discussion}\label{results}

As elementary pion production amplitude we use the MAID-2003
model which is parametrized in terms of invariant amplitudes allowing
the evaluation in any frame of reference. As in photoproduction, one
encounters the principal problem of off-shell continuation. In the
present work this problem is neglected by assuming on-shell kinematics
for the struck nucleon and the pion in the final state, because the
MAID-amplitudes do not allow an off-shell extrapolation. 
For the evaluation of the MAID amplitudes the invariant $\pi N$ energy
$W_{\pi N}$, the squared four momentum transfer $Q^2$, and the pion
angle $\theta_{\pi N}$ in the $\pi N$ c.m.\ system have to be 
specified. While $Q^2$ is given by the virtual photon, one has to
determine $W_{\pi N}$ and $\theta_{\pi N}$ from the kinematics of the
active nucleon in the $\gamma^*$-$d$ c.m.\ system. For this purpose we
assume as just mentioned that the four momenta $p_\pi$ and $p_f$ of
pion and active nucleon, respectively, in the final state obey the
on-shell condition. Then the needed $\pi N$ c.m.\ variables are
obtained by a Lorentz transformation with with boost parameter 
$\vec{\beta}=(\vec{p}_\pi+\vec{p}_f)/(E_\pi+E_f)$. The energy and
momentum of the initial off-shell nucleon then are determined through
the energy-momentum conservation at the elementary vertex, i.e.\
$p_i=p_\pi+p_f-q$. 

The explicit calculation of the $NN$-rescattering contribution follows the
same approach as in photoproduction~\cite{FiA05} by using the separable
representation of the realistic Paris potential from~\cite{HaP85} and
include all partial waves up to $^3D_3$. From previous results on
photoproduction it is expected that any realistic $NN$-potential model
will give very similar results. Thus the use of the Paris potential is
not crucial. Also with respect to the question whether the use of a
nonrelativistic $NN$-potential can be justified in view of the high
energies involved, we can refer to the remark in~\cite{FiA05}.
Similarly, we use for the evaluation of $\pi N$-rescattering a
realistic separable representation of the $\pi N$ interaction 
from~\cite{NoB90} and take into account all partial waves up to
$l=2$.

\subsection{Survey on semi-exclusive structure functions}
We will start with a general survey on the properties of the
semi-exclusive structure functions, unpolarized as well as polarized.  
With respect to the two possible charged pion channels we will
consider here $\pi^+$ production only, because the role of hadronic
FSI in $\pi^-$ production is expected to be very similar to the one in
$\pi^+$ production according to the results in
photoproduction~\cite{FiA05}.  
\begin{figure}[htb]
\includegraphics[scale=.55]{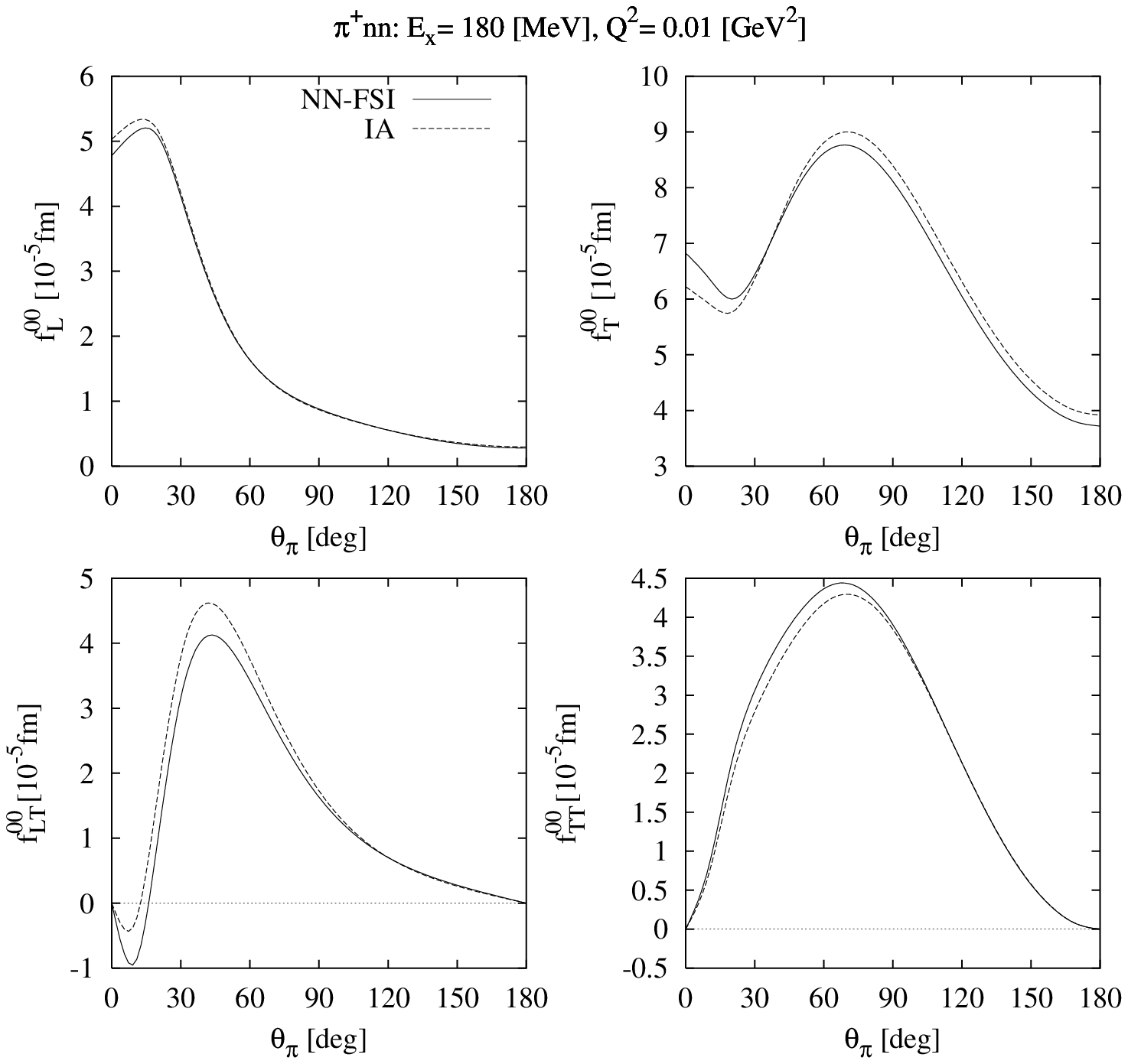}
\includegraphics[scale=.55]{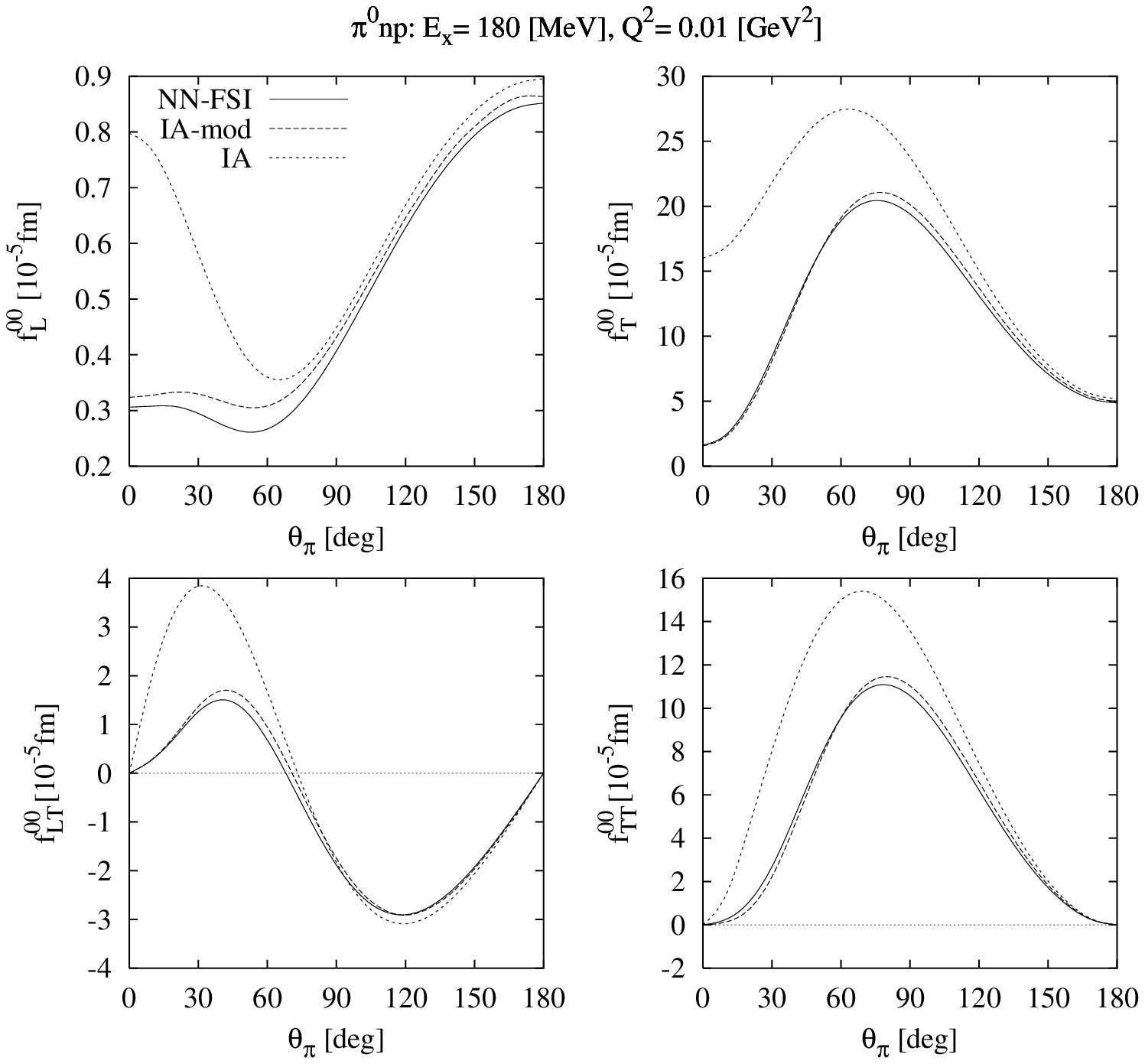}
\caption{Unpolarized structure functions for $\pi^+$ 
(left four panels) and $\pi^0$ electroproduction (right four panels)
at excitation energy $E_x=180$~MeV and squared four momentum
transfer $Q^2=0.01$~GeV$^2$ with $NN$-rescattering in the final state
(NN-FSI) and without (IA). For $\pi^0$ production results for the
modified IA are also given (IA-mod).} 
\label{fig_unpol}
\end{figure}

\subsubsection{Unpolarized semi-exclusive structure functions}

All four unpolarized structure functions for $\pi^+$ and $\pi^0$
electroproduction are shown in Fig.~\ref{fig_unpol} in IA and with 
inclusion of only $NN$-rescattering for an excitation energy
$E_x=180$~MeV ($E_x=W-2M-m_\pi$), which is in the region of the
$\Delta$-resonance, and for a quite low squared four-momentum transfer
$Q^2=0.01$~GeV$^2$. The dependence on $Q^2$ will be discussed later. 
The reason that we show only the influence of
$NN$-rescattering is that $\pi N$-rescattering is very small. This
fact has been noticed already by many
authors~\cite{Lag81,LeS00,DaA03,FiA05} for the case of incoherent
photoproduction of pions on the deuteron. Close to the threshold it
follows from the fact that the characteristic scale for $\pi N$-FSI
effects is given by the small ratio of the pion-nucleon scattering
length to the deuteron radius, i.e.\ by $a_{\pi N}/R_d \ll 1$. At
higher energies the insignificance of pion rescattering is related to
the smallness of the parameter $(pR_d)^{-1}$, where $p$ is a
characteristic momentum of the rescattered pion. As a consequence, 
the $\pi N$-interaction is much less effective in comparison to
$NN$-rescattering. 

One readily notes in Fig.~\ref{fig_unpol} that for charged pion
production the rescattering effects are in general quite small. It has
already been mentioned that they arise predominantly from
$NN$-rescattering while $\pi N$-rescattering is almost negligible. The
only exception is the near threshold region as will be shown below. 
For $\pi^+$ production $f_L^{00}$ exhibits a distinct forward
peak while $f_T^{00}$ possesses a much broader angular distribution
with a maximum around 75$^\circ$. In $f_L^{00}$ significant
FSI effects appear only at small angles below 30$^\circ$ where they
result in a small decrease. In contrast to this, one notes in
$f_T^{00}$ FSI effects over the whole angular range, in forward
direction a slight decrease and above 50$^\circ$ a small increase. 
The interference structure functions are of the same magnitude than
the diagonal ones. Both exhibit a maximum, around 40$^\circ$ for
$f_{LT}^{00}$ with a smaller widths and near 70$^\circ$ for
$f_{TT}^{00}$ with a broader distribution. FSI results in a slight reduction
in $f_{LT}^{00}$ and a small enhancement in $f_{TT}^{00}$. 

The unpolarized structure function for neutral pion production in
Fig.~\ref{fig_unpol} exhibit quite a different behavior. The influence
of FSI is dramatic which, however, stems predominantly from the
well-known fact, that in IA a large fraction of coherent production is
included because the final $NN$-plane wave is not orthogonal to the
deuteron bound state wave function. As is discussed in detail
in~\cite{FiA05} the effect of this non-orthogonality can be eliminated
by applying a modified IA where the deuteron wave function component
is projected out from the final $NN$-plane wave (see Appendix B
of~\cite{FiA05}). The additional influence of FSI then is indeed quite
small and comparable to charged pion production. In contrast to
$\pi^+$ production, $f_L^{00}$ exhibits a pronounced peak in backward
direction. However, in absolute size this structure function is much
smaller than $f_{T}^{00}$ and thus it is not surprising that FSI is
noteable over the whole angular region, particularly sizeable near the
minimum around 50$^\circ$. On the other hand, $f_{T}^{00}$ and the
interference structure functions $f_{LT}^{00}$ and $f_{TT}^{00}$ as
well show very little FSI effects. Interesting is the pronounced
forward-backward asymmetry of $f_{LT}^{00}$.
\begin{figure}[htb]
\includegraphics[scale=.55]{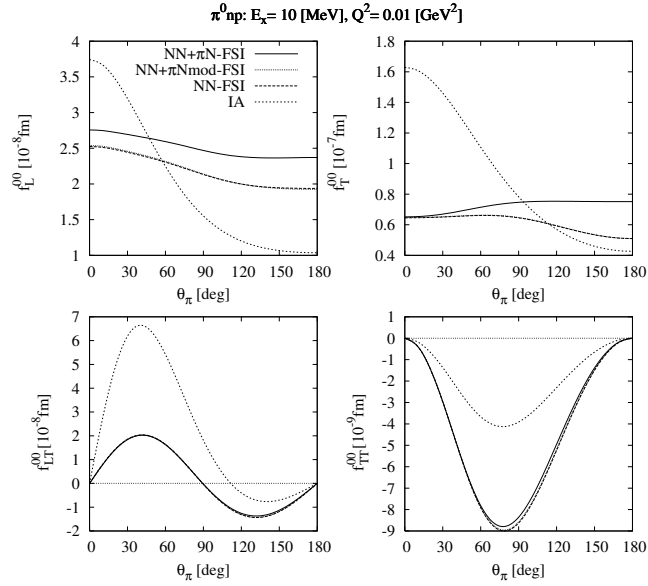}
\caption{Influence of $NN$ (NN-FSI) and additional $\pi N$
(NN+$\pi$N-FSI) final state rescattering compared to the impulse
approximation (IA) on unpolarized structure functions of $\pi^0$
electropoduction near threshold at excitation energy
$E_x=10$~MeV and squared four momentum transfer $Q^2=0.01$~GeV$^2$.
For the curve labeled ``NN-FSI+$\pi$Nmod'' the charge exchange 
contribution to the $\pi N$ FSI has been switched off.}
\label{fig_piN_10}
\end{figure}

As already mentioned, the $\pi N$-final state interaction plays a role
only near threshold. This is demonstrated in Fig.~\ref{fig_piN_10}
where we show for $\pi^0$ production the unpolarized structure
functions in the near threshold region, i.e.\ 10~MeV above
threshold. In particular, the diagonal structure functions $f_L^{00}$
and $f_T^{00}$ show quite a significant enhancement from $\pi
N$-rescattering. In contrast to this, the interference structure
functions are very little affected by the additional $\pi N$-FSI. But as
soon as the excitation energy approaches the first resonance region,
the influence of $\pi N$-FSI dies out rapidly. 
\begin{figure}[htb]
\includegraphics[scale=.55]{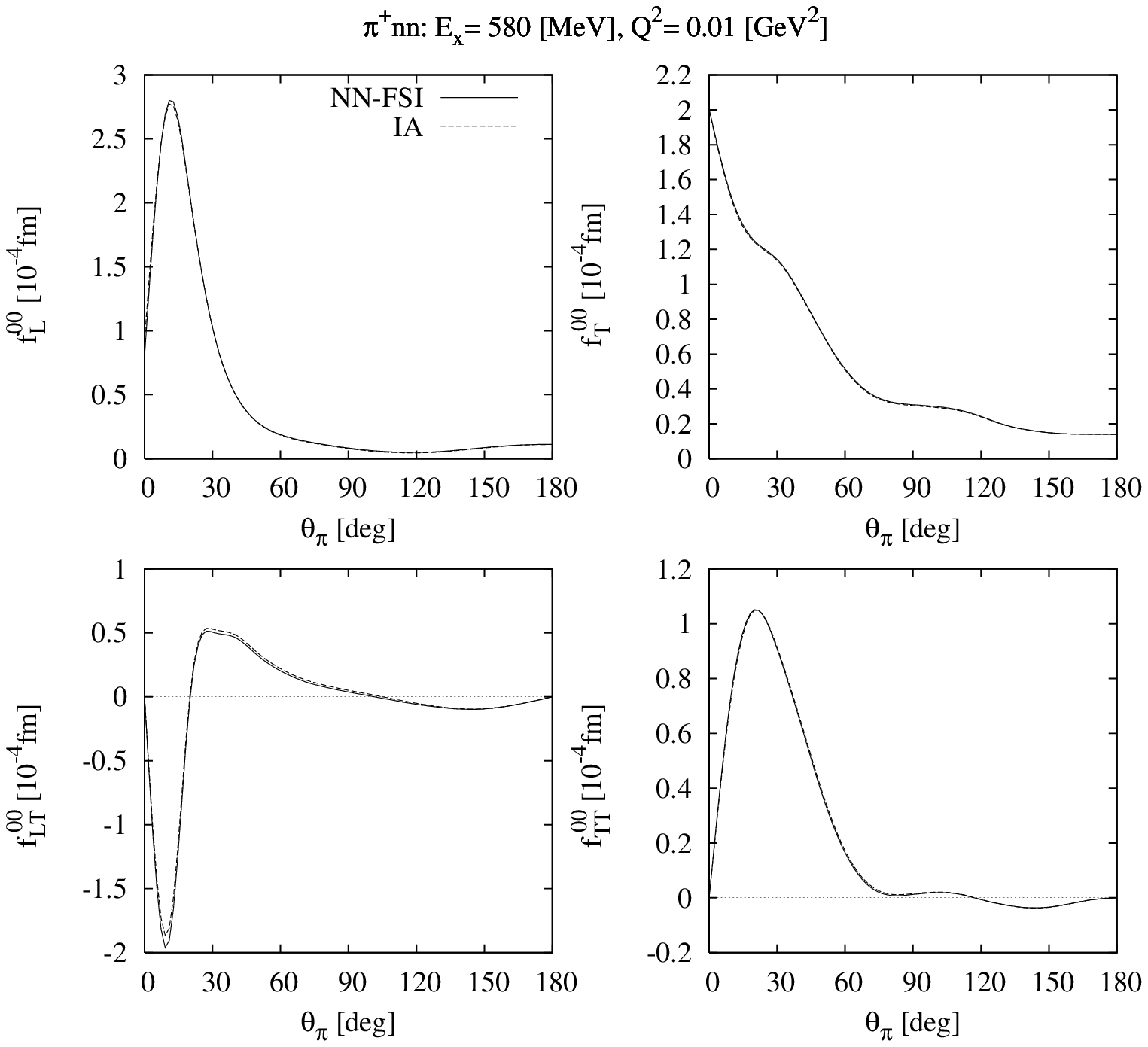}
\includegraphics[scale=.55]{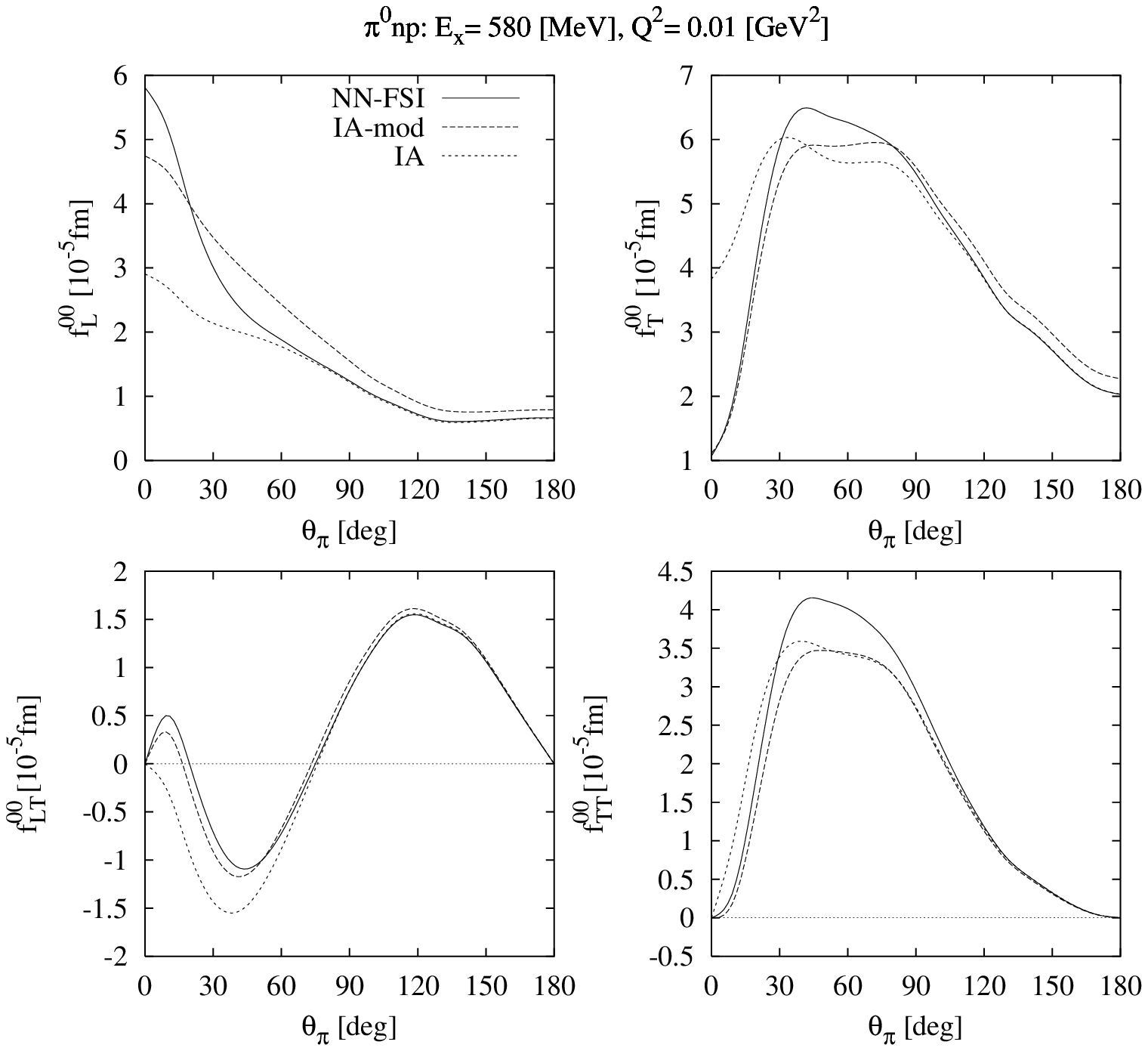}
\caption{Unpolarized structure functions for $\pi^+$ 
(left four panels) and $\pi^0$ electroproduction (right four panels)
at excitation energy $E_x=580$~MeV and squared four momentum
transfer $Q^2=0.01$~GeV$^2$ with $NN$-rescattering in the final state
(NN-FSI) and without (IA). For $\pi^0$ production results 
for the modified IA are also given.}
\label{fig_unpol_580_01}
\end{figure}
The sizeable influence of $\pi N$-FSI in $\pi^0$ production shown in
Fig.~\ref{fig_unpol} is mostly due to the strong suppression of the
IA cross section in the $\pi^0$ channel in the near threshold region.
As is well known, at very low kinetic energies of the active $\pi N$
system the dipole amplitude $E_{0+}$ in the neutral channel is about
an order of magnitude smaller than the one for $\pi^{\pm}$ production.
As a result, the dominant $\pi N$ rescattering effect in $\pi^0$
production is the charge-exchange mechanism, in which the production
of a charged pion on one of the nucleons is followed by a
charge-exchange rescattering on the second nucleon. This fact is
demonstrated in Fig.~\ref{fig_unpol} by the curve labeled ``NN-FSI+$\pi$Nmod'' 
representing the calculation for which this charge-exchange mechanism
has been switched off. 

The behavior of the unpolarized structure functions in the second
resonance region is displayed in
Fig.~\ref{fig_unpol_580_01}. One readily notes a significant change in
the angular distributions compared to the $\Delta$-regions. For
$\pi^+$ production, a more pronounced forward peaking is seen,
particularly for $f_{T}^{00}$ and $f_{TT}^{00}$. Final state
interaction effects
are almost negligible. This is not the case for $\pi^0$ production,
where FSI still has a significant influence, although much less than
in the $\Delta$-region. Also here a large fraction of FSI effects is
eliminated by the modified IA. The angular distributions are still
broad except for $f_{L}^{00}$ where one notes a forward increase. 
\begin{figure}[htb]
\includegraphics[scale=.55]{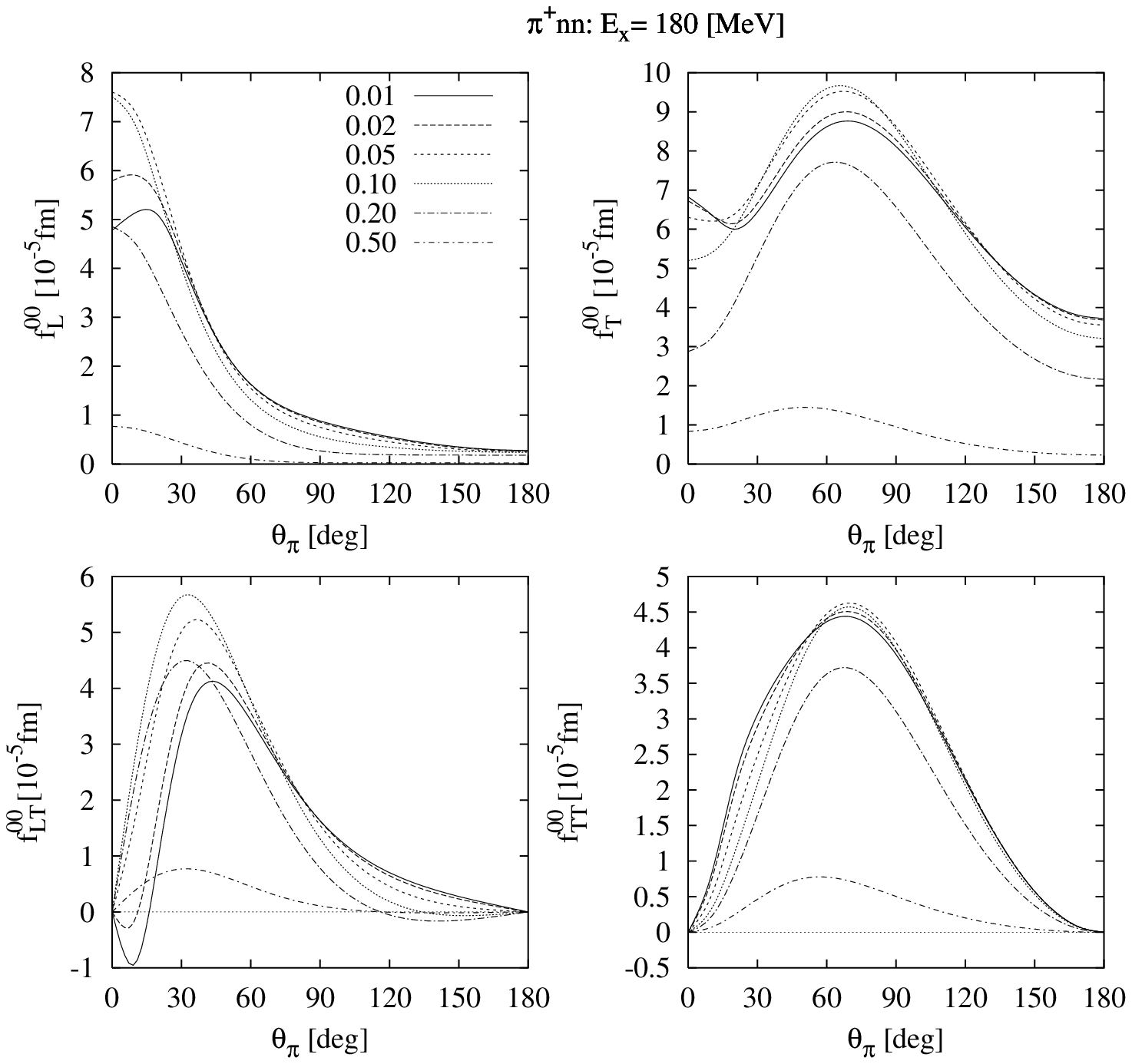}
\includegraphics[scale=.55]{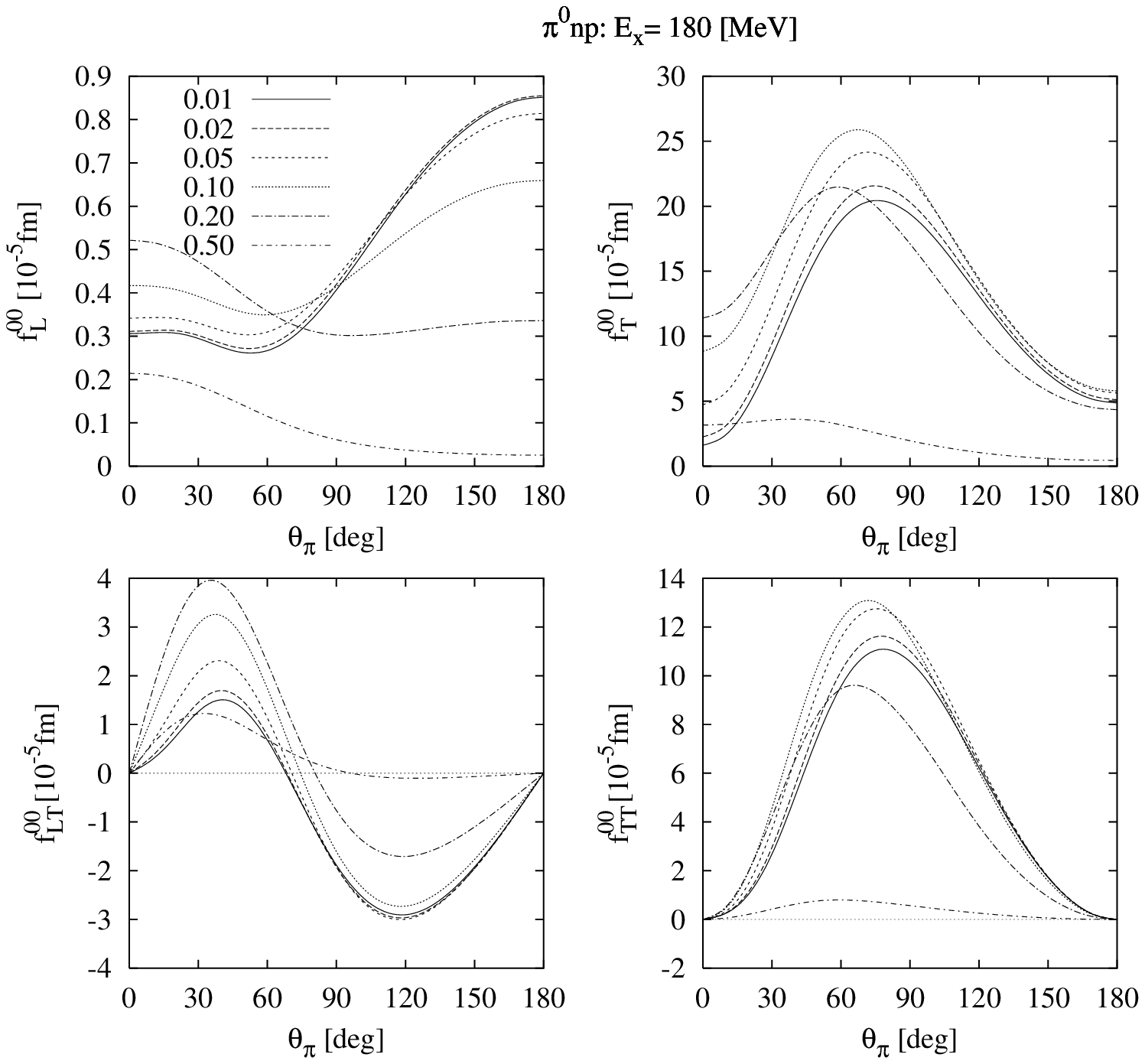}
\caption{$Q^2$-dependence of unpolarized structure functions for $\pi^+$ 
(left four panels) and $\pi^0$ electroproduction (right four panels)
at excitation energy $E_x=180$~MeV and various squared four
momentum transfers $Q^2=0.01,\,0.02,\,0.05,\,0.1,\,0.2$, and
0.5~GeV$^2$ as indicated in the legend with inclusion of
$NN$-rescattering in the final state.} 
\label{fig_unpol_E180}
\end{figure}

\begin{figure}[htb]
\includegraphics[scale=.55]{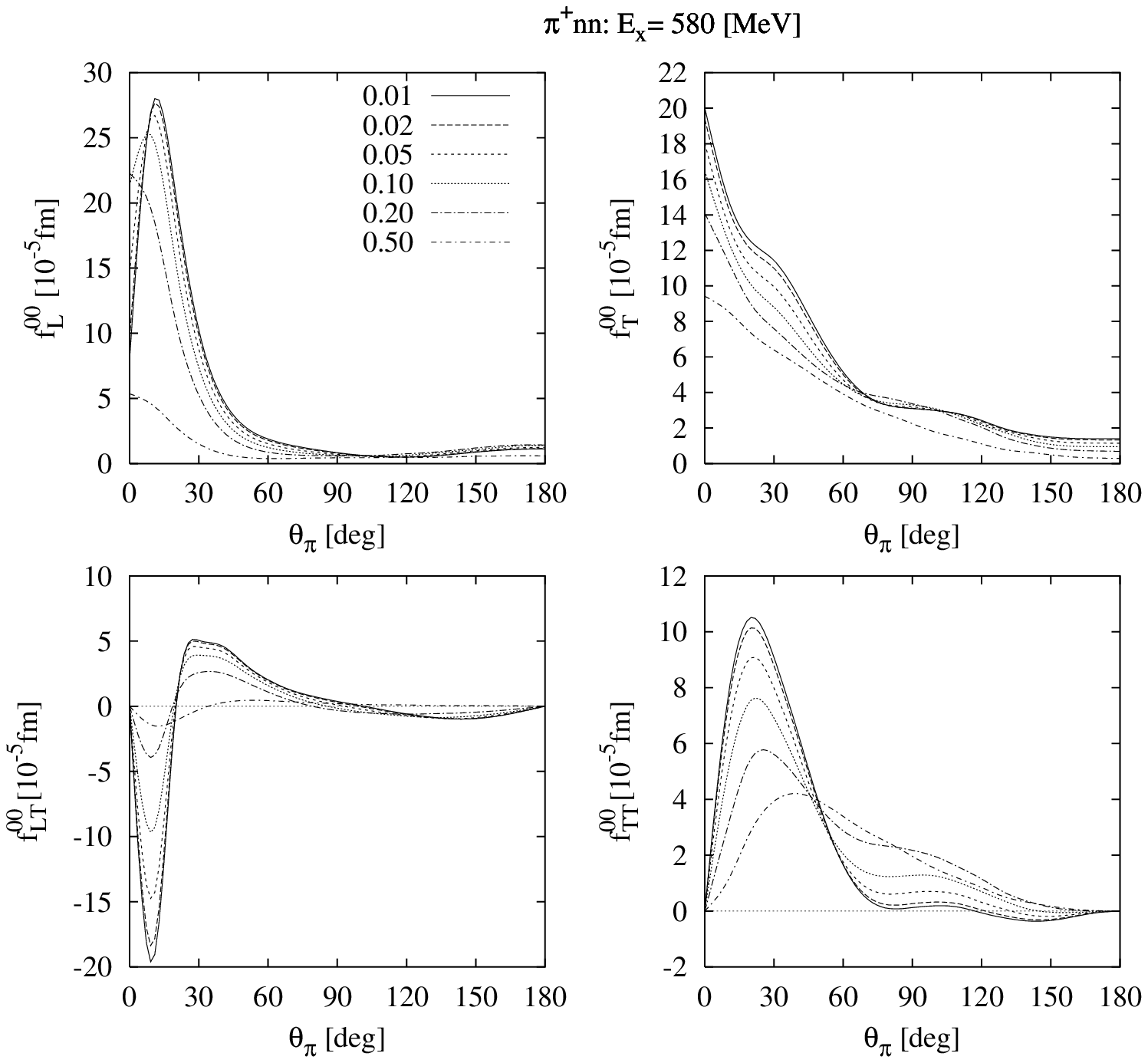}
\includegraphics[scale=.55]{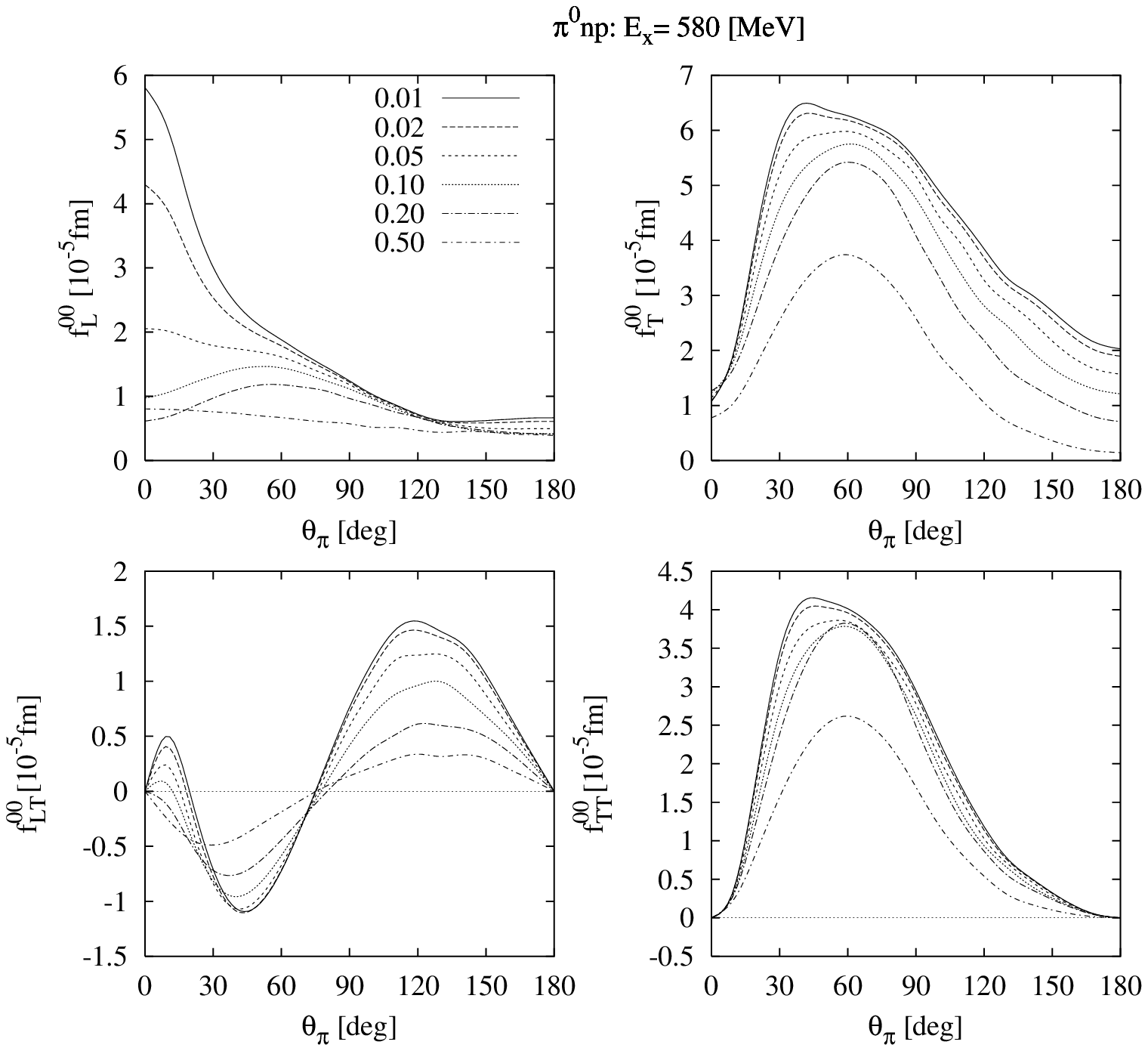}
\caption{As Fig.~\ref{fig_unpol_E180}
at excitation energy $E_x=580$~MeV.}
\label{fig_unpol_E580}
\end{figure}

The dependence on $Q^2$ is shown in Fig.~\ref{fig_unpol_E180} for
$E_x=180$~MeV on the $\Delta$-resonance and in
Fig.~\ref{fig_unpol_E580} for $E_x=580$~MeV in the second resonance
region. At both energies, the angular behavior of the structure
functions for different values of $Q^2$ 
look quite similar. For $\pi^+$ production the shape of the
four structure functions remains unchanged qualitatively except for
$f^{00}_{TT}$ at the lowest $Q^2$-value, only the size varies with
$Q^2$. In $f^{00}_{L}$ one first notes a slight increase going from
$Q^2=0.01$~GeV$^2$ to $Q^2=0.05$~GeV$^2$ but at higher $Q^2$ a steady rapid
decrease. The transverse structure function shows a general decrease
in size, whereas for $f^{00}_{LT}$ the decrease starts only after
$Q^2=0.05$~GeV$^2$. The transverse-transverse interference function
first increases slightly, then remains almost constant up to
$Q^2=0.1$~GeV$^2$ and finally becomes rapidly smaller at higher $Q^2$. 

In contrast to this, the structure functions for neutral pion
production in the right panels of Fig.~\ref{fig_unpol_E180} show quite
a different $Q^2$-dependence. Here, $f^{00}_{L}$ exhibits a strong
increase at forward angles with a corresponding decrease in backward
direction. For $f^{00}_{T}$ the maximum is slightly shifted towards
smaller angles while the amplitude remains constant up to
$Q^2=0.07$~GeV$^2$ and then falls off rapidly at higher $Q^2$. The
$LT$-interference function shows an increase of the forward maximum
and a decrease of the backward minimum. Only for $f^{00}_{TT}$ the
shape remains unchanged while the amplitude falls off with increasing
$Q^2$. Qualitatively, one finds a similar behavior at the higher
excitation energy $E_x=580$~MeV in the second resonance region
(Fig.~\ref{fig_unpol_E580}). 

\subsubsection{Polarized semi-exclusive structure functions for beam
and target polarization} 

For a longitudinally polarized electron beam and a polarized 
deuteron target the number of structure functions increases 
significantly. For that reason we show them in separate figures 
for longitudinal, transverse and the interference ones. 
\begin{figure}[htb]
\includegraphics[scale=.55]{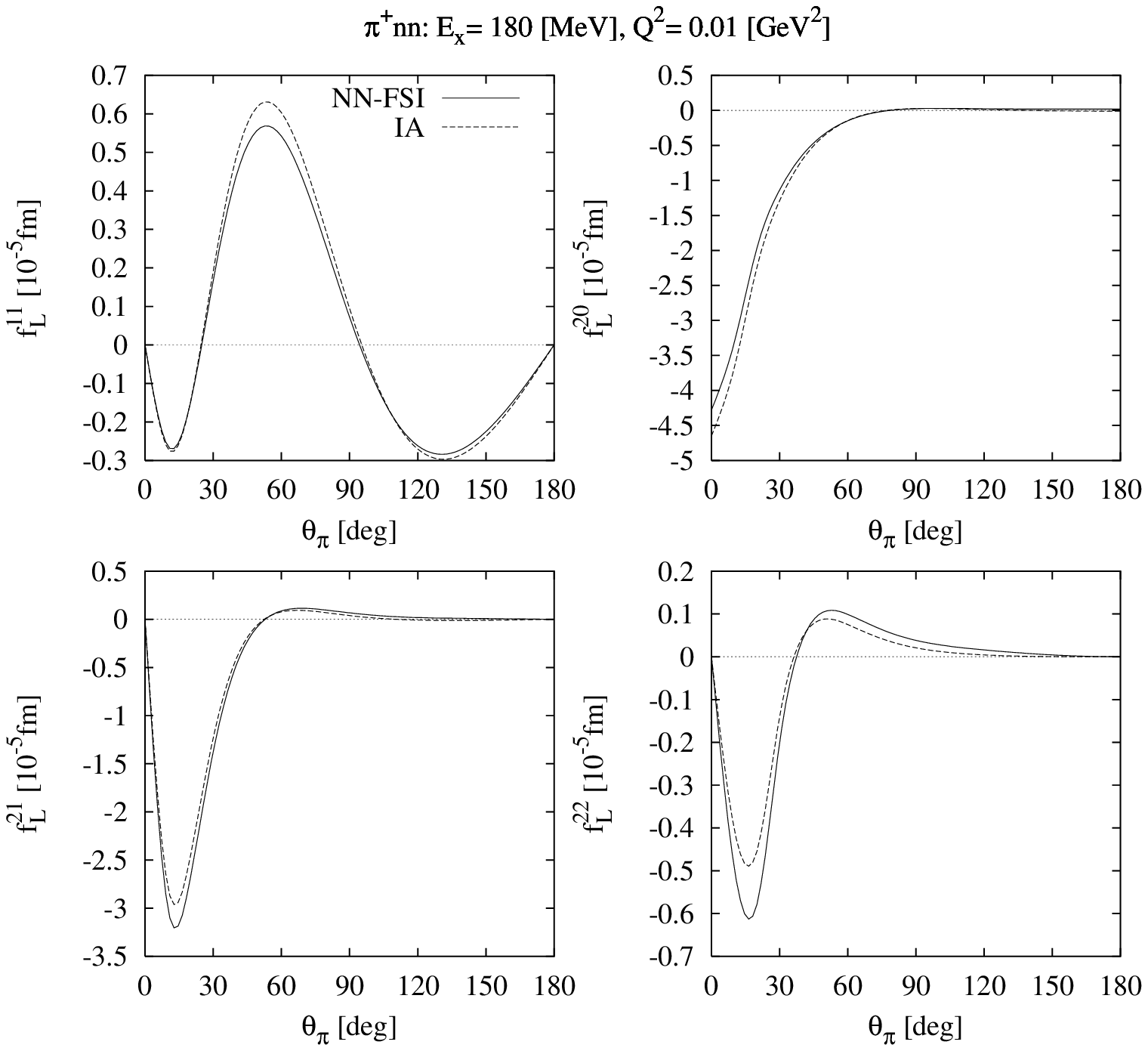}
\includegraphics[scale=.55]{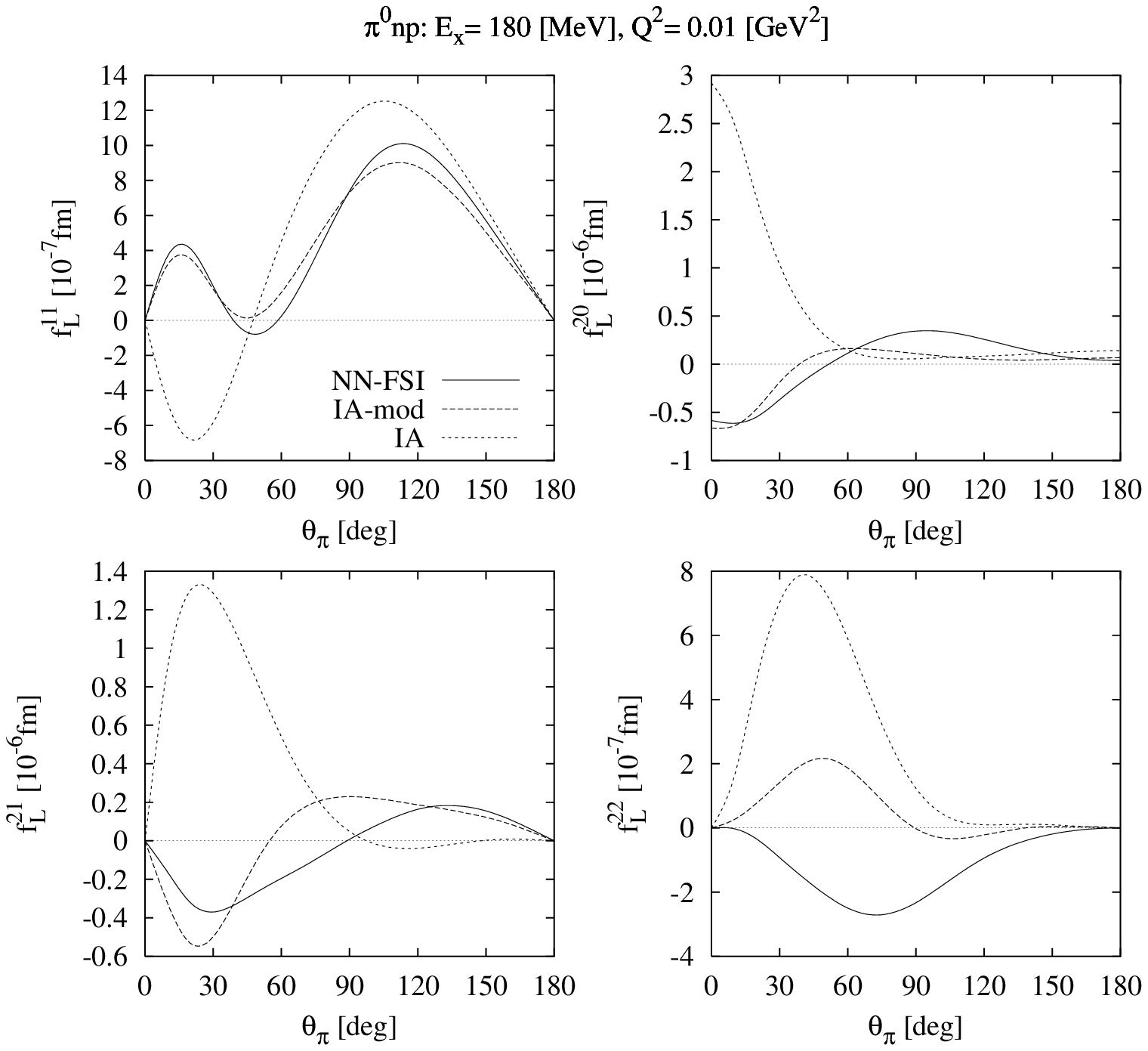}
\caption{Polarized longitudinal structure functions for $\pi^+$ 
(left four panels) and $\pi^0$ electroproduction (right four panels)
at excitation energy $E_x=180$~MeV and squared four momentum
transfer $Q^2=0.01$~GeV$^2$ with $NN$-rescattering in the final state
(NN-FSI) and without (IA). For $\pi^0$ production results 
for the modified IA are also given.}
\label{fig_pol_L}
\end{figure}
\begin{figure}[htb]
\includegraphics[scale=.55]{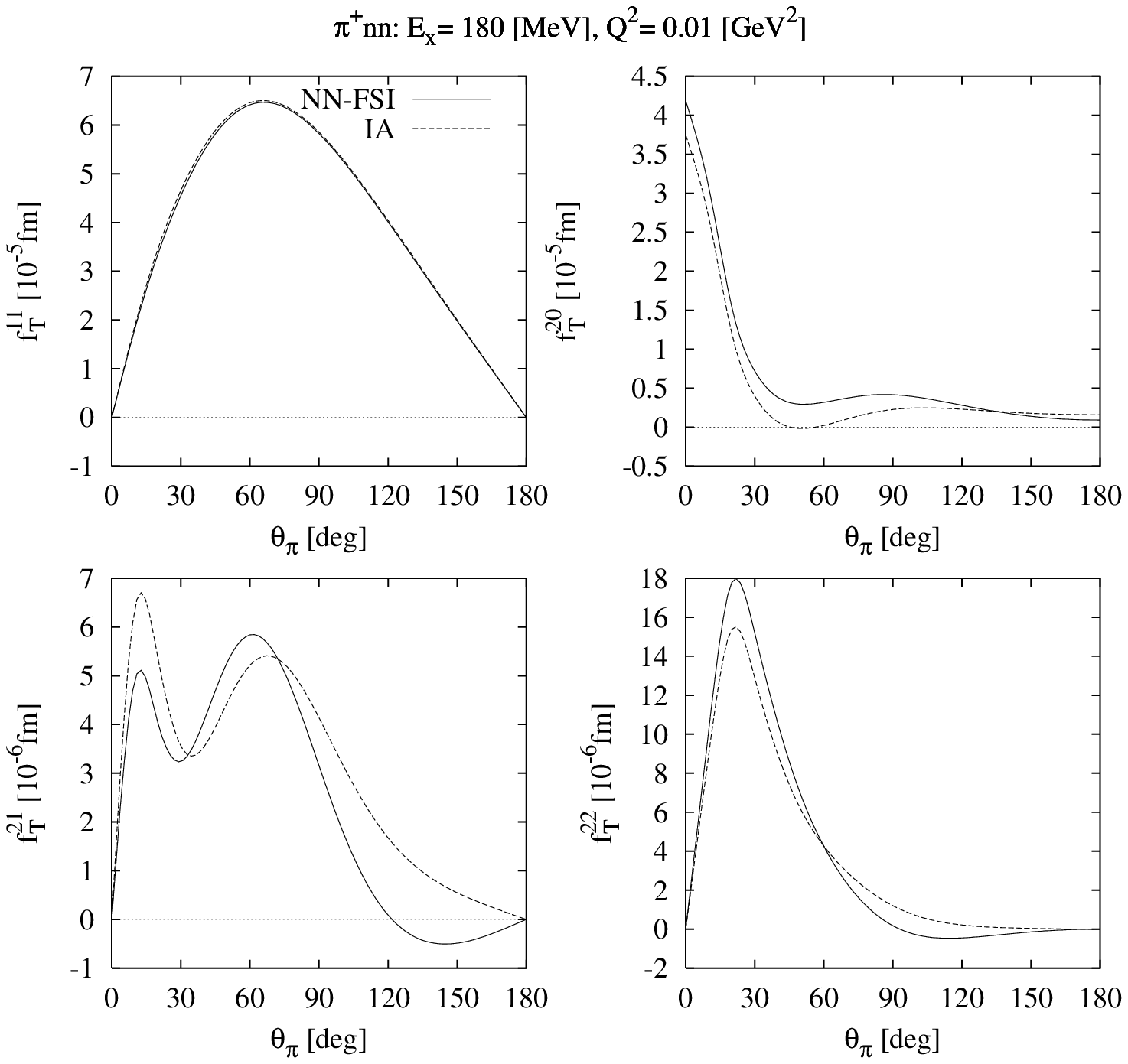}
\includegraphics[scale=.55]{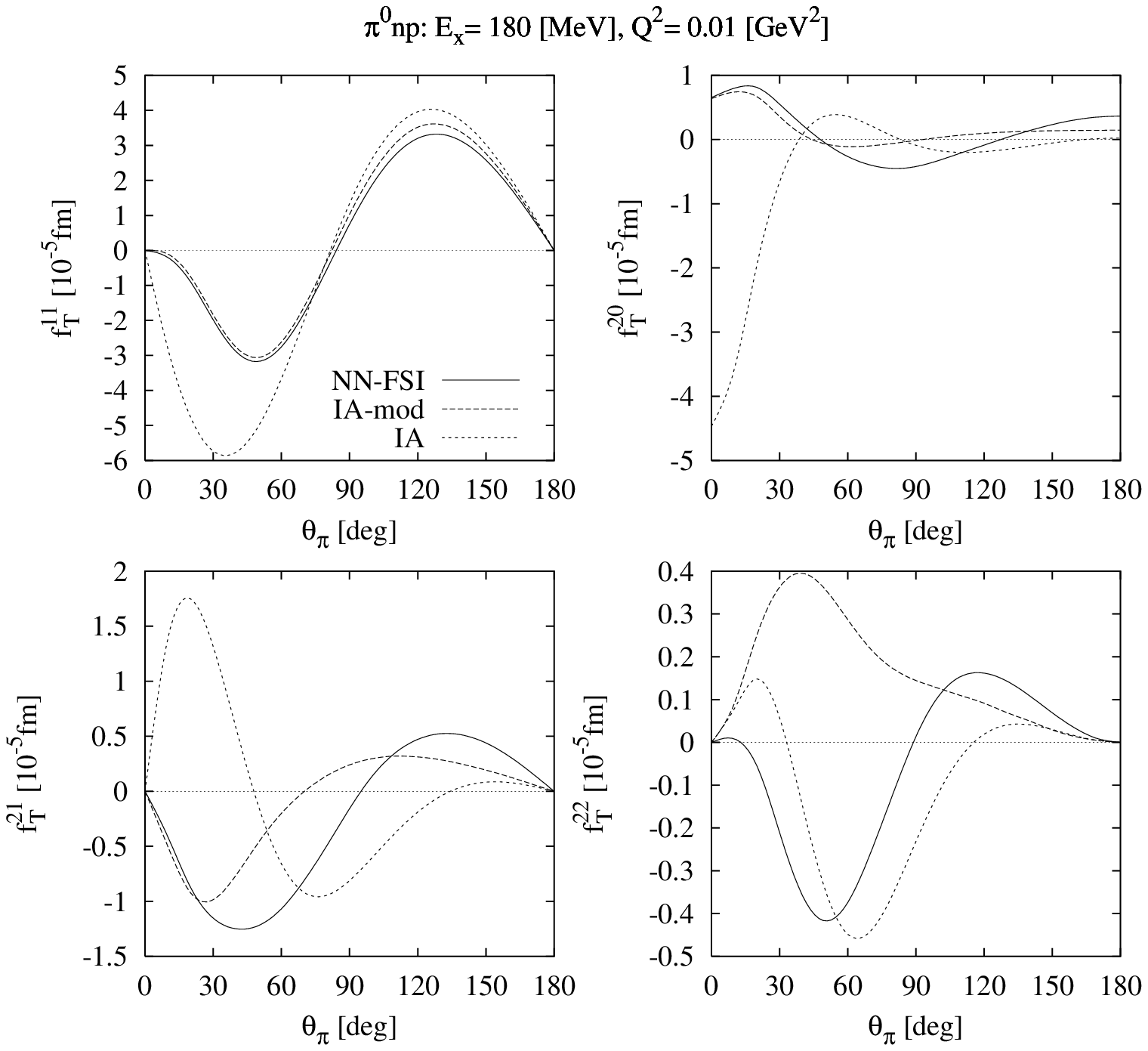}
\caption{Polarized transverse structure functions for $\pi^+$ 
(left four panels) and $\pi^0$ electroproduction (right four panels)
at excitation energy $E_x=180$~MeV and squared four momentum
transfer $Q^2=0.01$~GeV$^2$ with $NN$-rescattering in the final state
(NN-FSI) and without (IA). For $\pi^0$ production results 
for the modified IA are also given.}
\label{fig_pol_T}
\end{figure}
The polarized longitudinal structure functions in Fig.~\ref{fig_pol_L} 
for $\pi^+$ production exhibit in general a
pronounced forward peaking, except for $f_L^{11}$ showing an
oscillatory behavior with a maximum near 60$^\circ$. Since $f_L^{20}$
and $f_L^{21}$ are comparable in size to the unpolarized $f_L^{00}$,
one expects a sizeable dependence on the tesor polarization of an
oriented deuteron target. This has been pointed out already
in~\cite{LoP94}. The other two structure functions, $f_L^{11}$ and
$f_L^{22}$, are smaller by about a factor 5. For neutral pion
production the polarized structure functions are an order of magnitude
smaller. They exhibit a much broader angular distribution. FSI effects
are very strong as in the unpolarized case which, however, are again
largely reduced by using the modified IA. But the remaining FSI
effects are still quite significant, in particular for $f_L^{21}$ and
$f_L^{22}$. 
\begin{figure}[htb]
\includegraphics[scale=.5]{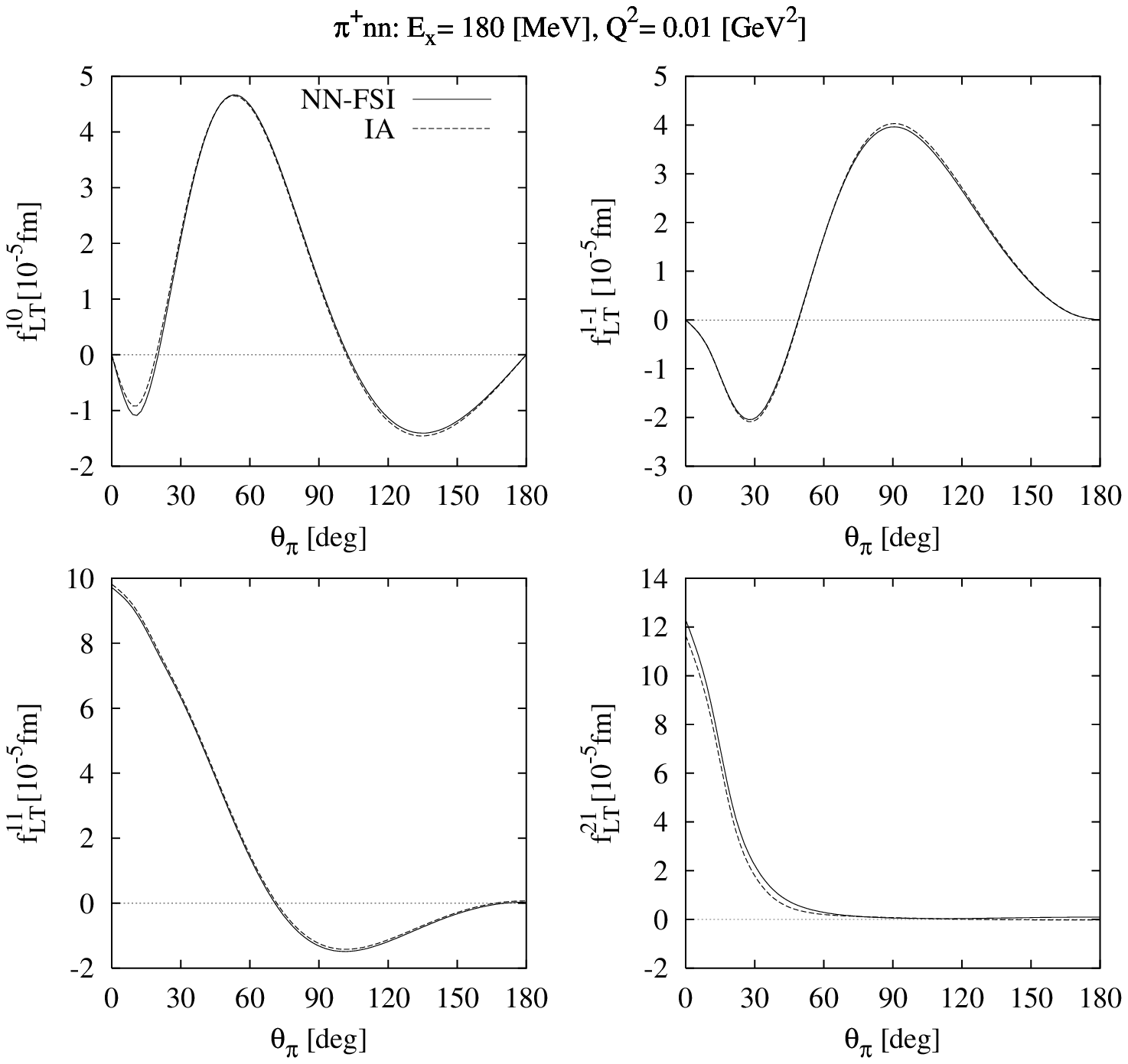}
\includegraphics[scale=.5]{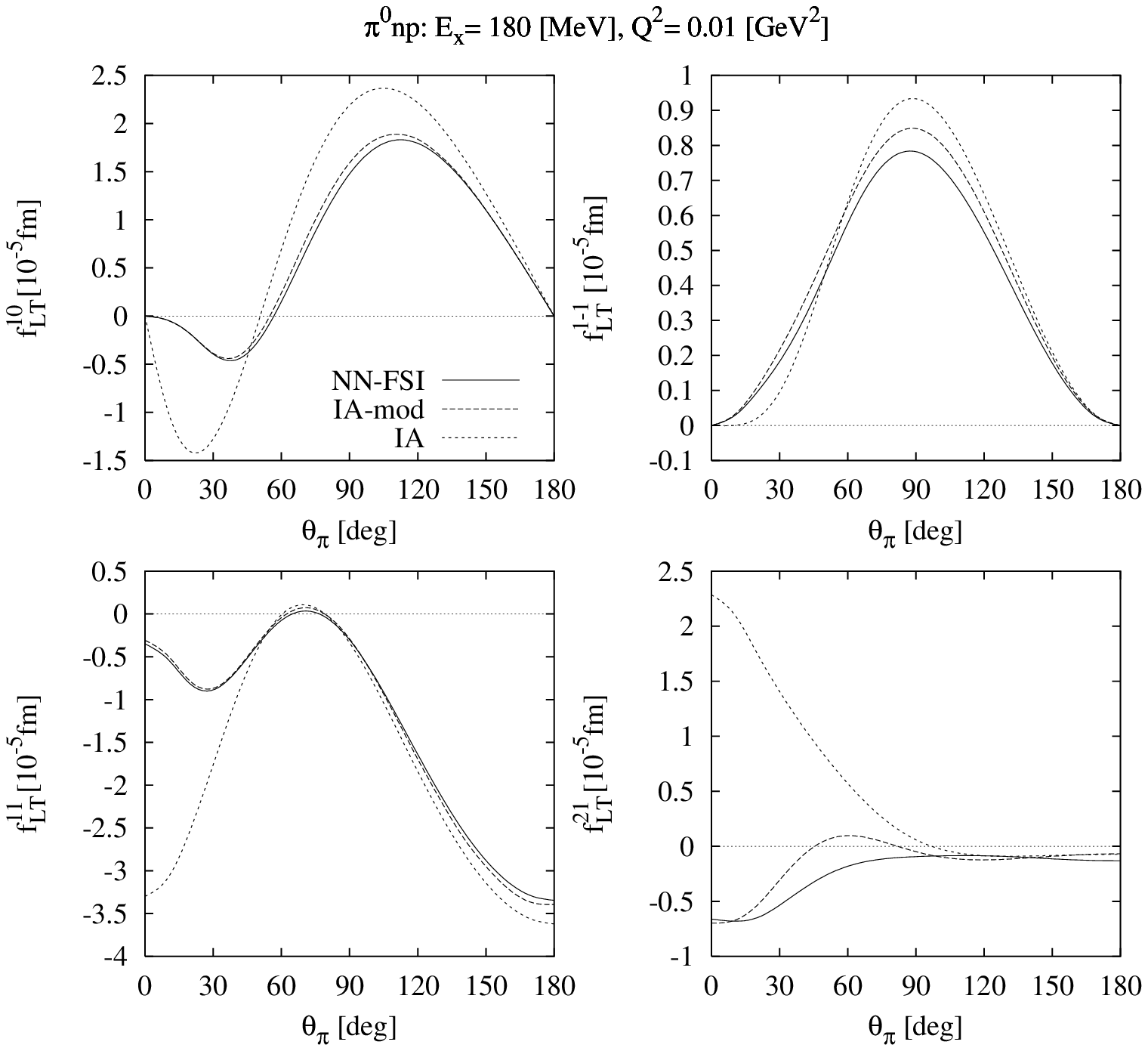}
\caption{Selected polarized longitudinal-transverse interference
structure functions for $\pi^+$ (left four panels) and $\pi^0$
electroproduction (right four panels) 
at excitation energy $E_x=180$~MeV and squared four momentum
transfer $Q^2=0.01$~GeV$^2$ with $NN$-rescattering in the final state
(NN-FSI) and without (IA). For $\pi^0$ production results 
for the modified IA are also given.}
\label{fig_pol_LT}
\end{figure}
\begin{figure}[htb]
\includegraphics[scale=.5]{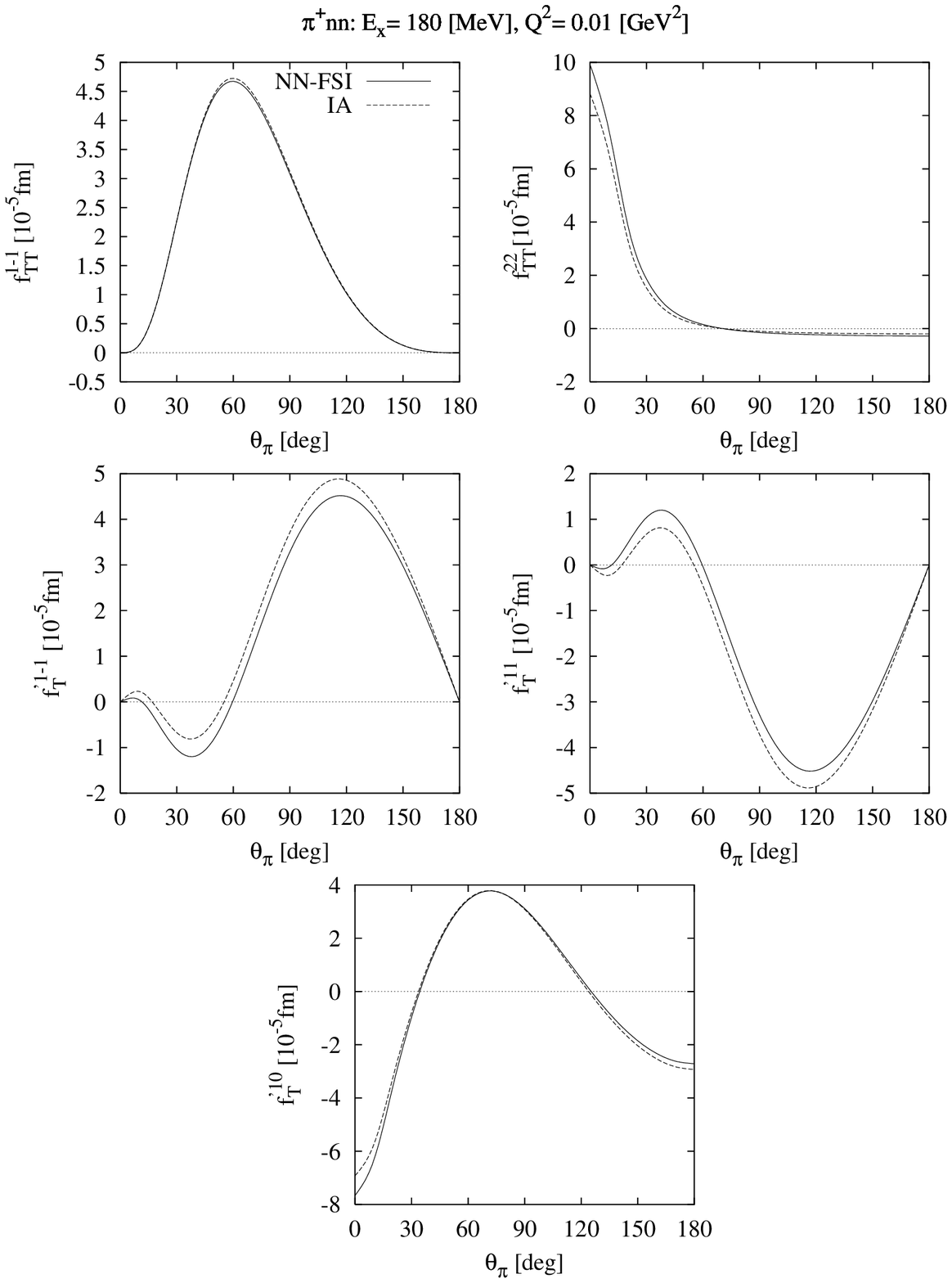}
\includegraphics[scale=.5]{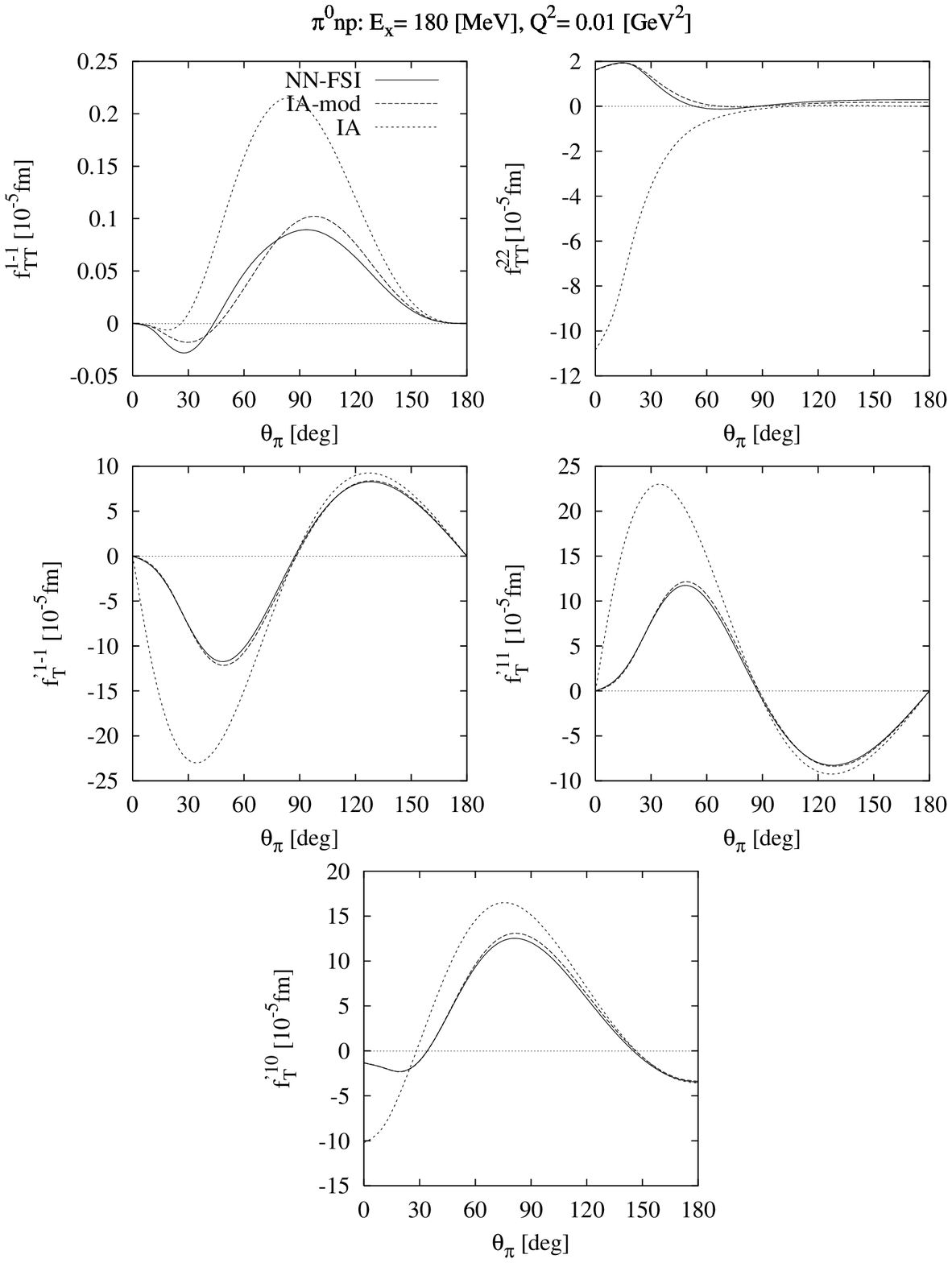}
\caption{Selected polarized structure functions of type $TT$ and for 
polarized electrons of type $T$ for $\pi^+$ 
(left five panels) and $\pi^0$ electroproduction (right five panels)
at excitation energy $E_x=180$~MeV and squared four momentum
transfer $Q^2=0.01$~GeV$^2$ with $NN$-rescattering in the final state
(NN-FSI) and without (IA). For $\pi^0$ production results 
for the modified IA are also given.}
\label{fig_pol_TTs}
\end{figure}

Much broader angular distributions show many of the transverse
structure functions in Fig.~\ref{fig_pol_T}. For $\pi^+$
production only $f_T^{20}$ and $f_T^{22}$ exhibit a strong forward
peaking. Comparable in size to the unpolarized $f_T^{00}$ is
$f_T^{11}$, and also $f_T^{20}$ is sizeable at small
angles. Remarkable is the fact that $f_T^{11}$ is almost independent
from FSI, whereas $f_T^{21}$ is quite sensitive to FSI. Compared to
$\pi^+$ one finds a different, more oscillatory angular
dependence. The largest one is $f_T^{11}$ while the other three
structure functions are considerably smaller. Additional FSI effects
beyond the modified IA are small in $f_T^{11}$ and $f_T^{20}$, but
more pronounced in the other structure functions of smaller absolute
size, in particular very strong in the smallest $f_T^{22}$. 

Of the eight polarized $LT$-interference structure functions we show
in Fig.~\ref{fig_pol_LT} the four most dominant ones which are
comparable in size to the unpolarized ones. They show quite different
characteristic angular behavior. In particular for $\pi^+$ production,
$f_{LT}^{11}$ and $f_{LT}^{21}$ possess a sizeable peak at
0$^\circ$. All structure functions are almost independent from FSI for
$\pi^+$ production, and even for $\pi^0$ production they exhibit
little FSI effects beyond the modified IA. 

\begin{figure}[htb]
\includegraphics[scale=.5]{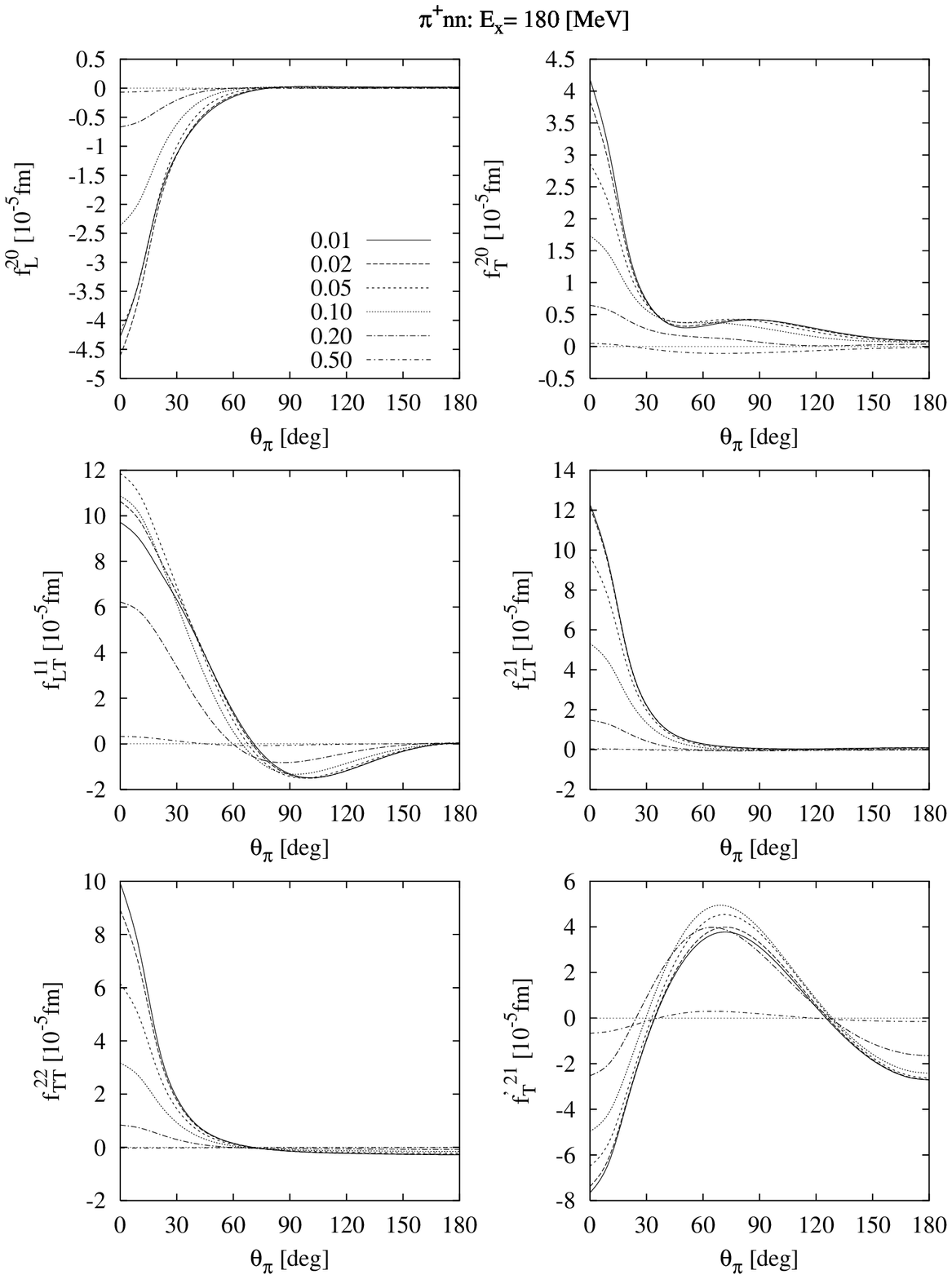}
\includegraphics[scale=.5]{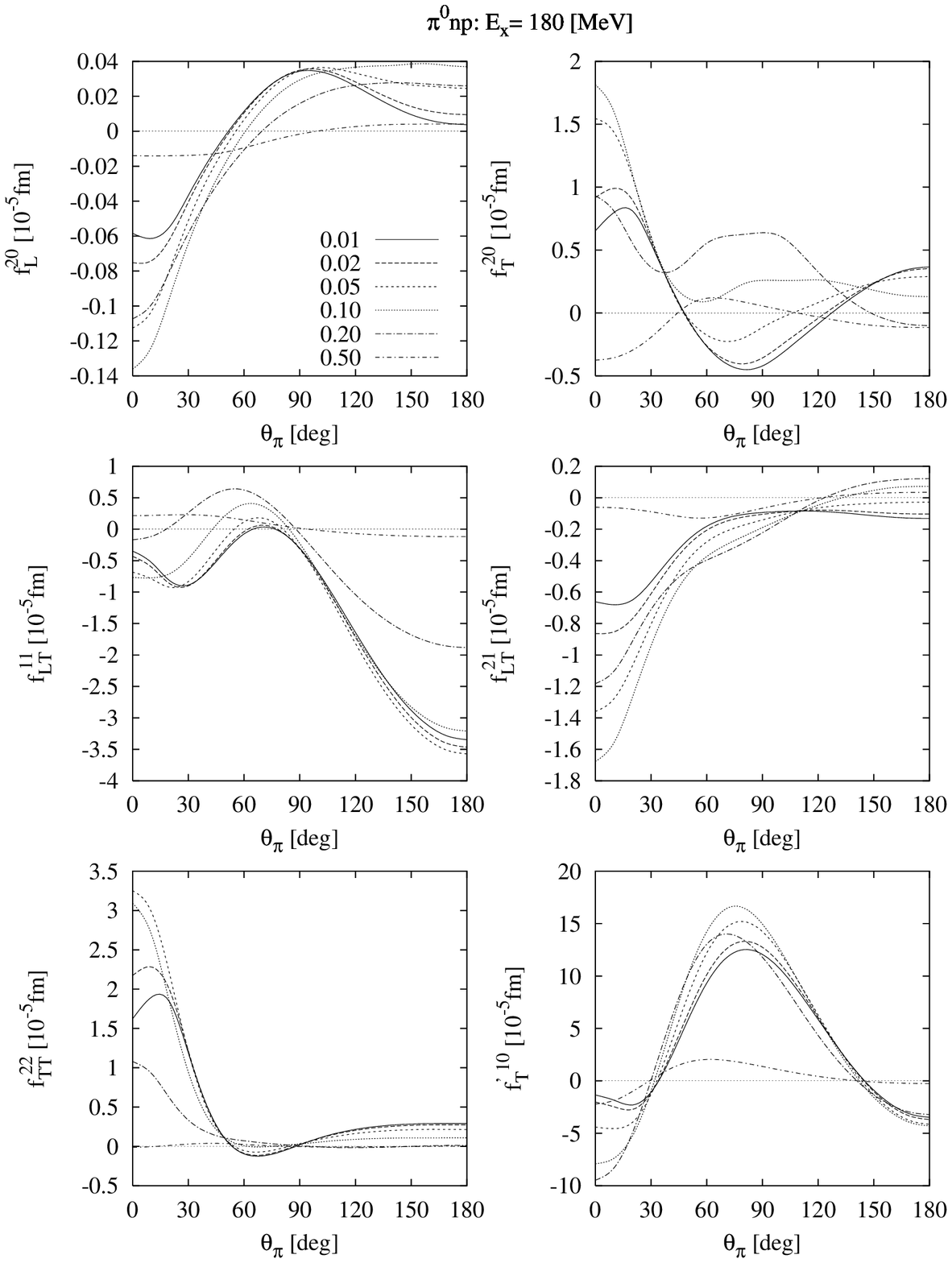}
\caption{$Q^2$-dependence of selected polarization structure functions
of $\pi^+$  (left six panels) and $\pi^0$ (right six panels)
electroproduction at excitation energy $E_x=180$~MeV and
various squared four momentum transfers
$Q^2=0.01,\,0.02,\,0.05,\,0.1,\,0.2$, and 0.5~GeV$^2$.} 
\label{fig_pol_180}
\end{figure}

Also of the remaining polarized structure functions of $TT$-type and
for longitudinally polarized electrons only a few are displayed in
Fig.~\ref{fig_pol_TTs}. Again we note very little influence of
$NN$-rescattering for both $\pi^+$ production as well as for $\pi^0$
production beyond the modified IA. Of particular interest is
$f_T^{\prime 10}$ which determines the contribution of single pion
production on the deuteron to the generalized
Gerasimov-Drell-Hearn sum rule~\cite{Are04}. 
To conclude this survey, we show in Fig.~\ref{fig_pol_180} the
$Q^2$-dependence of some selected examples of polarized structure
functions. 

\subsection{Comparison with other calculations and experiment}

\begin{figure}[htb]
\includegraphics[scale=.55]{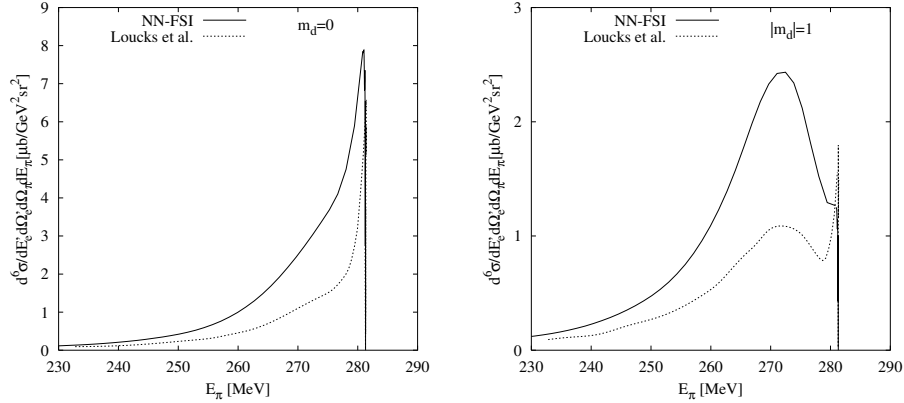}
\caption{Semi-exclusive differential cross sections for $\pi^+$
production on a deuteron with definite spin projection $m_d=0$ (left
panel) and $|m_d|=1$ (right panel) at $W=2126$~MeV for the kinematics
of the Saclay experiment~\cite{GiB90} ($E_e=645$~MeV,
$E_{e'}=355$~MeV, $\theta_e=36^\circ$). Notation: solid curves:
present results; dotted curves: results of Loucks et al.~\cite{LoP94}
divided by four.} 
\label{fig_loucks_fig4}
\end{figure}
\begin{figure}[htb]
\includegraphics[scale=.55]{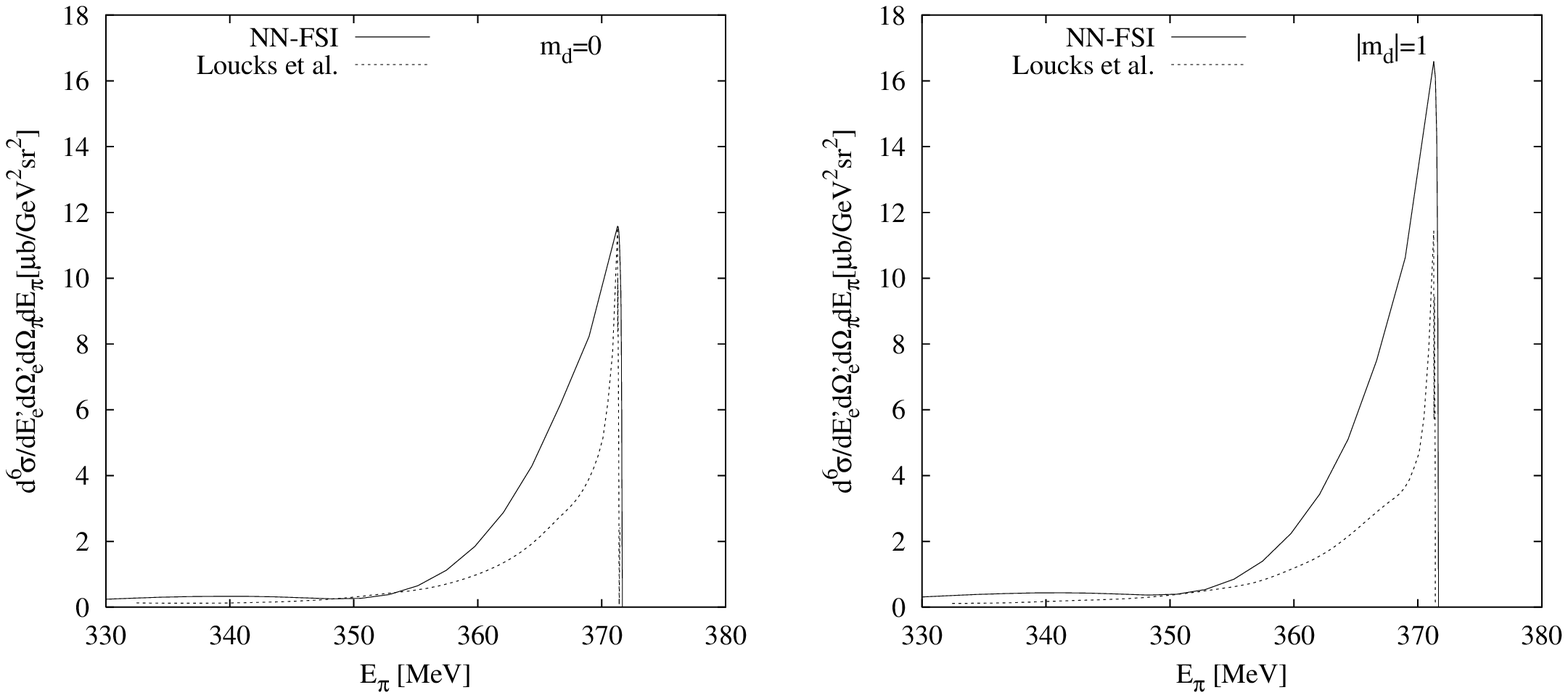}
\caption{Same as in Fig.~\ref{fig_loucks_fig4} at $W=2205$~MeV for the
kinematics of the Saclay experiment~\cite{GiB90} ($E_e=645$~MeV,
$E_{e'}=269$~MeV, $\theta_e=36^\circ$).} 
\label{fig_loucks_fig5}
\end{figure}
We will begin the comparison with an earlier calculation by Loucks et
al.~\cite{LoP94} who have studied the influence of FSI for the
kinematics of the Saclay experiment~\cite{GiB90} using a simple pion
production model. In particular, they found a strong dependence of the
semi-exclusive cross section on the deuteron orientation by
considering the idealized case of an intial deuteron state being
prepared in a state 
with definite spin projection $m_d^0$ on the momentum transfer. This
means in our formalism deuteron orientation angles
$\theta_d=0$ and $\phi_d=0$ and a deuteron density matrix $\rho_{m_d\,
{m_d}'}^d=\delta_{m_d\, {m_d}'}\delta_{m_d\, m_d^0}$. The
corresponding vector and tensor orientation parameters $P_1^d$ and
$P_2^d$, respectively, as function of $m_d^0$ are
obtained from (\ref{rhodpar}). Specifically, one finds
\beq
\renewcommand{\arraystretch}{1.8}
\begin{array}{lll}
m_d^0=0: & P_1^d(0)=0\,,&P_2^d(0)=-\sqrt{2}\,,\cr
m_d^0=\pm 1: & P_1^d(\pm 1)=\pm \frac{3}{\sqrt{2}}\,,
&P_2^d(\pm 1)=\frac{1}{\sqrt{2}}\,.\cr
\end{array}
\eeq
In fact, these two values for $P_2^d$ mark the maximal and minimal
possible values for the tensor polarization.
For these two cases, one finds from (\ref{diffcross_semi_inclusive})
for pion emission along $\vec q$
\beqa
\frac{d^6\sigma(\Omega_\pi=(0,0),\Omega_d=(0,0))}
{dE'_e d\Omega'_e dp_\pi d\Omega_\pi}\Big|_{m_d=0}&=&
\frac{\alpha_{qed}}{Q^4}\,\frac{k_e'}{k_e}\Big[
\rho_L\widetilde f_L^{00}+\rho_T\widetilde f_T^{00}
-\sqrt{2}(\rho_L\widetilde f_L^{20}+\rho_T\widetilde f_T^{20})\Big]\,,\\
\frac{d^6\sigma(\Omega_\pi=(0,0),\Omega_d=(0,0))}
{dE'_e d\Omega'_e dp_\pi d\Omega_\pi}\Big|_{m_d=\pm 1}&=&
\frac{\alpha_{qed}}{Q^4}\,\frac{k_e'}{k_e}\Big[
\rho_L\widetilde f_L^{00}+\rho_T\widetilde f_T^{00}
+\frac{1}{\sqrt{2}}(\rho_L\widetilde f_L^{20}
+\rho_T\widetilde f_T^{20})\Big]\,.
\eeqa
We compare in Figs.~\ref{fig_loucks_fig4} and \ref{fig_loucks_fig5}
the results of~~\cite{LoP94} with the present calculation. In view of
the fact, that 
the elementary production model of~\cite{LoP94} gives a cross section
too large by about a factor four compared to experiment as pointed out
in~\cite{LeC04}, we have renormalized the results of~\cite{LoP94} by
this factor. Qualitatively, we find the same dependence on the
orientation of the deuteron, namely, for $m_d=0$ a dominant
contribution from the antibound $^1S_0$-$NN$-state near the
$NN$-threshold at $W=2126$~MeV masking completely the quasi-free peak
which is located at $E_\pi^{qf,lab}=271.8$~MeV according to
(\ref{Epiquasifreelab}) (left panel of 
Fig.~\ref{fig_loucks_fig4}), whereas for $|m_d|=1$ one notes a strong
suppression of the $^1S_0$-state at the same invarient energy $W$ so
that the quasi-free peak becomes clearly seen (right panel of
Fig.~\ref{fig_loucks_fig4}). At the higher energy $W=2205$~MeV in
Fig.~\ref{fig_loucks_fig5} the dependence on the orientation 
is much weaker. For both orientations the $^1S_0$-peak is the dominant
feature, and the quasi-free peak at $E_\pi^{qf,lab}=340.5$~MeV is barely seen.
However, on a quantitative level, we find quite
significant differences to the results of~\cite{LoP94} apart from the
overall strength. At $W=2126$~MeV, we obtain a much stronger
suppression of the $^1S_0$-state appearing only as a
shoulder. Furthermore, at lower pion energies, i.e. with increasing
$NN$-energy, we find significantly higher strength near the quasi-free
peak by almost a factor two. The origin of this difference is not
clear. Also at the higher energy $W=2205$~MeV one notes again much
higher strengths at lower pion energies and moreover also a slightly
stronger dependence on the tensor polarization. 
\begin{figure}[htb]
\includegraphics[scale=.55]{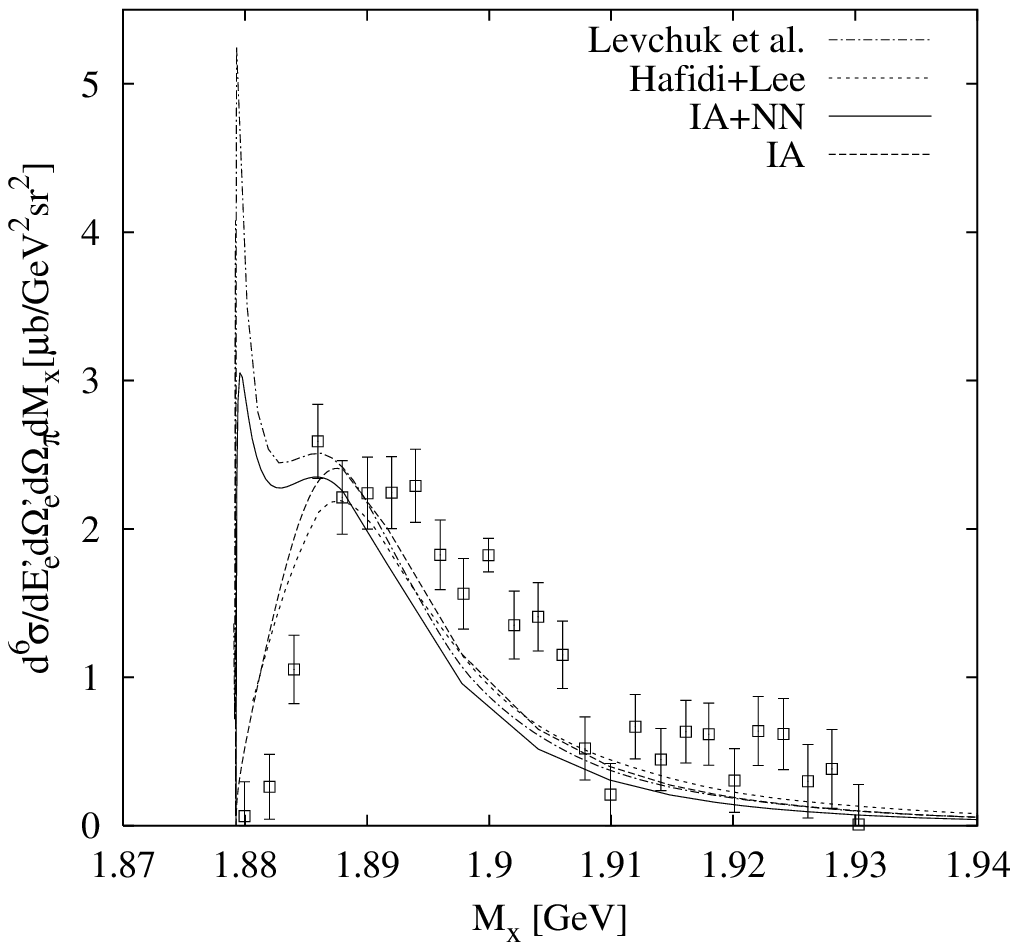}
\includegraphics[scale=.55]{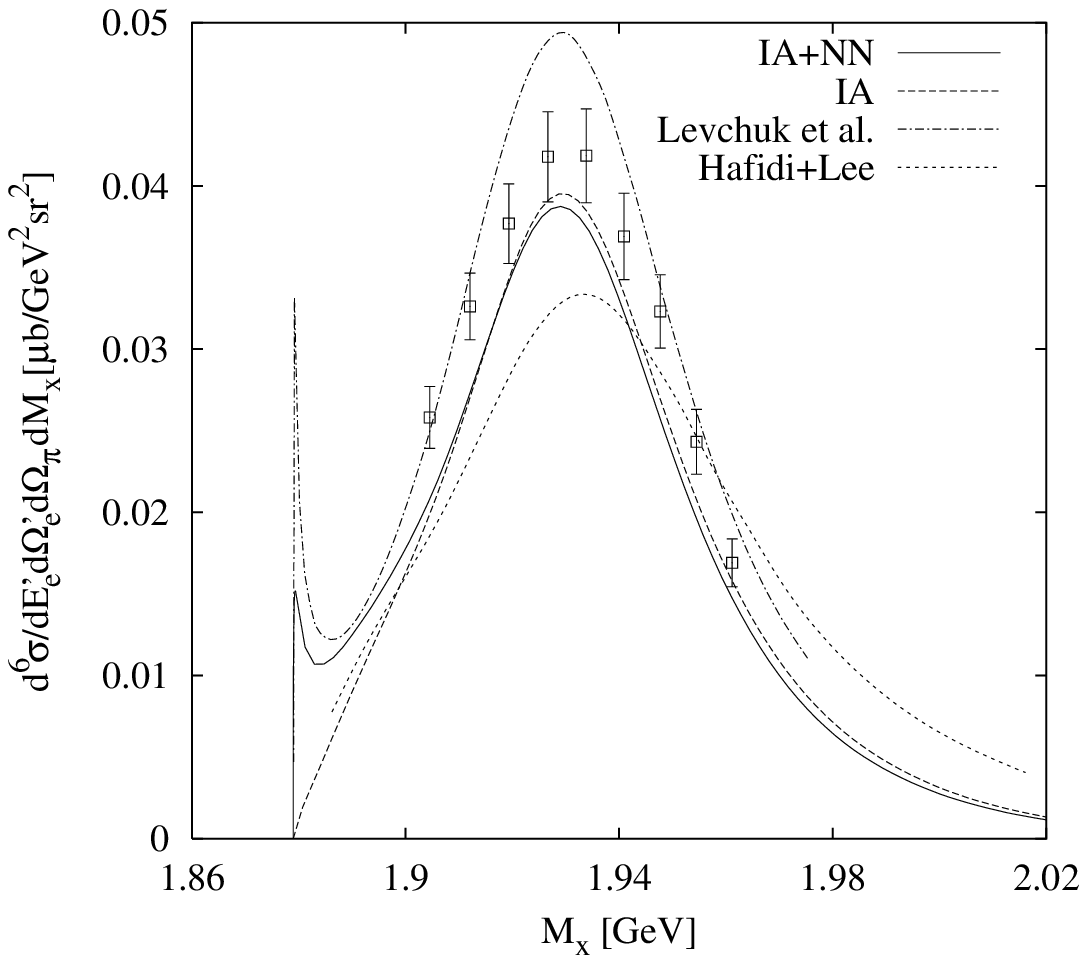}
\caption{Semi-exclusive differential cross sections for $d(e,e'\pi^+)$
in parallel kinematics, i.e.\ pion emission along the momentum
transfer as functions of the missing mass $M_x$. Left panel:
kinematics of Saclay data~\cite{GiB90} 
($E_e=645$~MeV, $E_e'=355$~MeV, $\theta_e'=36^\circ$); 
Right panel: kinematics of Jefferson Lab data~\cite{GaA01}
($E_e=845$~MeV, $E_e'=395$~MeV, $Q^2=0.4$~GeV$^2$). Notation 
of curves: solid: present calculations with NN-FSI included; dashed:
IA alone; dash-dot: calculations of Hafidi and Lee~\cite{HaL01};
dotted: calculations of Levchuk et al.~\cite{LeC04}.} 
\label{fig_exp}
\end{figure}
 Now we turn to a comparison of our results with experimental data and
with the calculations of Hafidi and Lee~\cite{HaL01} and Levchuk et
al.~\cite{LeC04} in Fig.~\ref{fig_exp}. The left panel shows the
semi-exclusive differential cross section for pion emission in
forward direction for the kinematics of the Saclay data~\cite{GiB90}
while the right panel shows the one for the kinematics of the
Jefferson Lab data~\cite{GaA01}. 

For the Saclay data (left panel of Fig.~\ref{fig_exp}) all three
calculations give very similar results for the maximum of the data,
which coincides with the position of the quasi-free kinematics, i.e.\
$M_x^{qf}=1885$~MeV, and
at higher missing mass all of them slightly underestimate the
data. However, at lower missing mass our calculation 
as well as the one of Levchuk et al.\ exhibit the sharp and very
pronounced $^1S_0$-peak right at threshold which is absent in the
calculation of Hafidi and Lee and also not seen in the data. It
remains a puzzle why this pronounced peak, which is seen in deuteron
electrodisintegration near threshold~\cite{SiW79} as well as in pion
photoproduction~\cite{KoA87}, is absent in the data of~\cite{GiB90}. 

In contrast to this, the three theoretical results for the kinematics
of the Jefferson Lab data (right panel of Fig.~\ref{fig_exp}) differ
substantially in the quasi-free maximum ($M_x^{qf}=1931$~MeV). While
the results of Levchuk et al.\ overshoot the 
maximum, the one of Hafidi and Lee underestimate it by about the same
amount. In this case, our calculation gives a fairly good account of
the data although one still notes a slight systematic
underestimation. Near threshold we and also Levchuk et al.\ find again
a sharp $^1S_0$-peak but differing in height while in the results of
Hafidi and Lee there appears no indication of such a peak at all. 

\section{Conclusion and outlook}\label{conclusions}

In the present paper we have investigated the influence of final
state interactions in pion electroproduction on the deuteron for
energies from threshold up to the second resonance region and squared
momentum transfers between 0.01~GeV$^2$ and 0.5~GeV$^2$. Special
emphasis was laid on the study of polarization observables for beam
and target polarization. Formal expressions for all structure
functions as quadratic hermitean forms in the production amplitudes
were derived which govern the unpolarized differential cross 
section as well as the various polarization observables extending thus
the formal developments for pion photoproduction in~\cite{ArF05}.

The semi-exclusive structure functions of $d(e,e'\pi)NN$ were then
evaluated taking the elementary 
operator for $eN\to e'\pi N'$ from the MAID-2003 analysis. The
interaction in the final state was included on the basis of the
two-body $t$-matrices for $NN$ and $\pi N$ scattering, respectively,
restricting their contribution to the first order in the multiple scattering
series. With respect to the energy region above the $\Delta(1232)$
resonance, the present work represents the first realistic calculation
which extends into the second resonance region. The results show that
primarily the $NN$-interaction is important leading to sizeable
modifications of some of the structure functions, especially in the
energy region up to the $\Delta$ resonance. On the other hand,
the effect of pion rescattering is significant and should be included
only close to threshold, much below the $\Delta$ resonance. At higher
energies it can safely be neglected. 

Most visibly distorted by the interaction between the final particles 
are the unpolarized as well as polarized structure functions in the
$\pi^0$ channel in contrast to the $\pi^+$ channel which is much less
sensitive to FSI effects. As in the case of photoproduction, this
strong effect in the $\pi^0$ channel is almost completely due to a
spurious contribution of the coherent reaction because of the
non-orthogonality of the deuteron wave function to the plane wave of
the final two nucleons in IA. After removing this contribution, the
remaining FSI effect is similar in size to what has been found in the
charged channel. 

With respect to the few existing experimental data we find in general
a satisfactory agreement with the data for the missing mass spectra of
$d(e,e'\pi^+)$ in parallel kinematics measured at Saclay~\cite{GiB90}
and JLab~\cite{GaA01} around the quasi-free peaks.
In comparison to the theoretical results of~\cite{LoP94} we find a
similar strong dependence on the tensor polarization of an oriented
deuteron target in the near threshold region. However, we obtain a much
stronger quasi-free peak relative to the $^1S_0$ spike right at
threshold. With respect to the work of~\cite{HaL01} we see a much
stronger influence of the $NN$ interaction at very low excitation
energies in the $NN$ subsystem as manifest by the $^1S_0$ peak which
is not present in the results of~\cite{HaL01}. On the other hand, we
do not find this peak so pronounced as predicted in~\cite{LeC04}. The
origin of these differences is not clear. 

In view of the fact that at present only few data are available for
the unpolarized semi-exclusive differential cross section of $\pi^+$
electroproduction at very forward angles, there is an urgent need for
more measurements of angular distributions at higher energies and for
various momentum transfers for all three charge states of single pion
electroproduction on the deuteron. Furthermore, polarization data are
totally missing, although they would provide a more detailed analysis
of this reaction, in particular, a more detailed investigation of the
elementary reaction on the neutron in $\pi^-$ production on the
deuteron. It is therefore very desirable to have new precise
experiments in order to improve our knowledge of pion
electroproduction on nucleon and deuteron. 

\appendix*
\renewcommand{\theequation}{A\arabic{equation}}
\setcounter{equation}{0}
\section{Relation to other formal expressions}
\label{appa}

In this appendix we would like to give the relations of the structure
functions $\widetilde f_\alpha^{00}$ defined in (\ref{strufu}) to 
other ones used in
the literature. We begin with the expression for the semi-exclusive
differential cross section of Loucks et al.~\cite{LoP94}
\beqa
\frac{d^6\sigma}{dE_{e'} d\Omega_{e'} dE_\pi d\Omega_\pi}&=&
\sigma_M\frac{p_\pi E_\pi M p}{12(2\pi)^3}\Big[v_L R_L+v_T R_T 
+v_{LT}R_{LT}-v_{TT} R_{TT} \Big]\,,\label{loucksetal}
\eeqa
where we assume that all quantities refer to the lab system although
this is not stated explicitly in~\cite{LoP94}. In fact, in this
appendix all variables refer to the lab system if not stated otherwise
and indicated specifically. Here, the Mott cross section is denoted by
$\sigma_M=\alpha_{qed}^2\cos^2(\theta_e/2)/4E_e^2\sin^4(\theta_e/2)$, 
and the kinematic functions $v_\alpha$ with $\alpha\in\{L,T,LT,TT\}$
are related to the virtual photon density matrix in (\ref{rhos}) by 
\beq
v_\alpha=\frac{2\eta}{Q^2}\beta^{-g_\alpha}(\sqrt{2})^{\delta_{\alpha,LT}}
(-)^{\delta_{\alpha,TT}}\rho_\alpha\,, 
\eeq
where we have introduced for convenience 
\beq
g_\alpha =2\delta_{\alpha,L}+\delta_{\alpha,LT}\,.
\eeq
Comparing (\ref{loucksetal}) with (\ref{diffcross_semi_inclusive}),
one finds the following relation
\beqa
R_\alpha(E_\pi,\Omega_\pi)&=&
(-\frac{1}{\sqrt{2}}\cos\phi_\pi^{c.m.})^{\delta_{\alpha,LT}} 
(-\cos 2\phi_\pi^{c.m.})^{\delta_{\alpha,TT}}\beta^{g_\alpha}
\frac{6(2\pi)^3}{\alpha_{qed}Mp\,p_\pi^2}\nonumber\\
&&\times J(p_\pi^{c.m.},\Omega_\pi^{c.m.};E_\pi,\Omega_\pi)\,
\widetilde f_\alpha^{00}(p_\pi^{c.m.},\theta_\pi^{c.m.})
\,,
\eeqa
where the Jacobian 
\beq
J(p_\pi^{c.m.},\Omega_\pi^{c.m.};E_\pi,\Omega_\pi)=
\Big|\frac{\partial(p_\pi^{c.m.},\Omega_\pi^{c.m.})}
{\partial(E_\pi,\Omega_\pi)}\Big|=\frac{p_\pi E_\pi^{c.m.}}{(p_\pi^{c.m.})^2}
\eeq
takes care of the transformation from the c.m.\ frame variables to the
lab frame ones, because we had defined the structure functions
$f_\alpha^{00}$ with respect to c.m.\ variables. 

As next we consider the expression for the semi-exclusive
differential cross section of Levchuk et al.~\cite{LeC04} 
\beqa
\frac{d^6\sigma}{dE_{e'} d\Omega_{e'} dE_\pi d\Omega_\pi}&=&
\sigma_M p_\pi E_\pi\Big[\xi^2 W_C+(\eta+ \frac{\xi}{2})W_T 
-\xi\sqrt{\eta +\xi}\cos\phi_\pi\,W_I+\frac{\xi}{2}\cos 2\phi_\pi\,W_S\Big]\,,
\label{levchuketal}
\eeqa
where $\eta$ and $\xi$ are given in (\ref{betaxieta}). Noting the
relations to the virtual photon density matrix in (\ref{rhos})
\beq
\xi^2=\frac{1}{\beta^2}\frac{2\eta}{Q^2}\rho_L\,,\quad
\eta+ \frac{\xi}{2}= \frac{2\eta}{Q^2}\rho_T\,,\quad 
\xi\sqrt{\eta +\xi}=\frac{\sqrt{2}}{\beta}\frac{2\eta}{Q^2}\rho_{LT}\,,\quad
\frac{\xi}{2}=-\frac{2\eta}{Q^2}\rho_{TT}
\eeq
with $\beta$ also given in (\ref{betaxieta}),
and changing slightly the notation by setting $W_L=W_C$ and 
$W_{LT}=W_I$, one finds as relation
\beq
W_\alpha(E_\pi,\theta_\pi)=(-)^{\delta_{\alpha,TT}}
(-\frac{1}{\sqrt{2}})^{\delta_{\alpha,LT}} 
\frac{\beta^{g_\alpha}}{2\alpha_{qed}\,p_\pi^2}\,
J(p_\pi^{c.m.},\Omega_\pi^{c.m.};E_\pi,\Omega_\pi)\,
\widetilde f_\alpha^{00}(p_\pi^{c.m.},\theta_\pi^{c.m.})\,.
\eeq

Finally, we will consider the parametrization in terms of a 
virtual photon flux times a virtual photon cross section as, 
for example, used in~\cite{HaL01,GaA01}, i.e.\
\beq
\frac{d^6\sigma}{dE_{e'} d\Omega_{e'} dM_x d\Omega_\pi}=
\Gamma\,\frac{d^3\sigma_v}{dM_x d\Omega_\pi}\,,
\eeq
where $M_x=\sqrt{W^2+m_\pi^2-2WE_\pi^{c.m.}}$ denotes the 
missing mass. The virtual photon flux $\Gamma$ is defined by
\beq
\Gamma=\frac{\alpha_{qed}}{2\pi^2}\,\frac{E_{e'}}{E_e}\,
\frac{K}{Q^2}\,\frac{1}{1-\varepsilon}\,,
\eeq
where $K=(W^2-M_d^2)/2 M_d$ in~\cite{HaL01} and $K=(W^2-M^2)/2 M$
in~\cite{GaA01}, and the virtual photon cross section by
\beqa
\frac{d^3\sigma_v}{dM_x d\Omega_\pi}=
\frac{d^3\sigma_T}{dM_x d\Omega_\pi}
+\varepsilon\frac{d^3\sigma_L}{dM_x d\Omega_\pi}
+\sqrt{2\varepsilon(1+\varepsilon)}
\frac{d^3\sigma_{LT}}{dM_x d\Omega_\pi}\cos\phi_\pi
+\varepsilon\frac{d^3\sigma_{TT}}{dM_x d\Omega_\pi}\cos2\phi_\pi\,,
\eeqa
where $\varepsilon=\xi/(\xi+2\eta)$. In~\cite{HaL01} only the first
two terms were included since only pion emission along $\vec q$ was
considered where the the two interference terms vanish. Using the
relations 
\beq
\frac{1}{1-\varepsilon}=\frac{2}{Q^2}\rho_T\,,\quad
\varepsilon=-\frac{\rho_{TT}}{\rho_T}=
\frac{1}{2\beta^2\xi}\,\frac{\rho_L}{\rho_T}
\,,\quad  \sqrt{2\varepsilon(1+\varepsilon)}=
\sqrt{\frac{2}{\beta^2\xi}}\frac{\rho_{LT}}{\rho_T}\,,
\eeq
one finds for the cross section a form analogous to our expression 
in (\ref{diffcross_semi_inclusive})
\beqa
\frac{d^6\sigma}{dE_{e'} d\Omega_{e'} dM_x d\Omega_\pi}&=&
\frac{\alpha_{qed}}{\pi^2}\,\frac{E_{e'}}{E_e}\,
\frac{K}{Q^4}
\Big[\frac{1}{2\beta^2\xi}\rho_L\frac{d^3\sigma_L}{dM_x d\Omega_\pi}
+\rho_T\,\frac{d^3\sigma_T}{dM_x d\Omega_\pi}\nonumber\\
&&+\sqrt{\frac{2}{\xi}}\frac{1}{\beta}\rho_{LT}
\frac{d^3\sigma_{LT}}{dM_x d\Omega_\pi}\cos\phi_\pi
-\rho_{TT}\,\frac{d^3\sigma_{TT}}{dM_x d\Omega_\pi}\cos2\phi_\pi\Big].
\eeqa
Comparison of this expression with (\ref{diffcross_semi_inclusive})
yields as final relation
\beq
\frac{d^3\sigma_\alpha}{dM_x d\Omega_\pi}=
\frac{(\beta\sqrt{2\xi})^{g_\alpha}}{2^{\delta_{\alpha,LT}}}\,
\frac{\pi^2}{K}\,
J(p_\pi^{c.m.},\Omega_\pi^{c.m.};M_x,\Omega_\pi)\,
\widetilde f_\alpha^{00}(p_\pi^{c.m.},\theta_\pi^{c.m.})\,.
\eeq
with
\beq
J(p_\pi^{c.m.},\Omega_\pi^{c.m.};M_x,\Omega_\pi)=
\Big(\frac{p_\pi}{p_\pi^{c.m.}}\Big)^2
\frac{E_\pi^{c.m.}M_x}{|(q_0+M_d)p_\pi -E_\pi q\cos\theta_\pi|}
\,.
\eeq

\end{document}